%% file: main.tex
\documentclass[a4paper,onecolumn,11pt,unpublished]{quantumarticle}
\pdfoutput=1

\usepackage[utf8]{inputenc}
\usepackage[english]{babel}
\usepackage[T1]{fontenc}
\usepackage{amsmath,amssymb}
\usepackage{booktabs}
\usepackage{tabularx}
\usepackage{xltabular}
\usepackage{tikz}
\usetikzlibrary{arrows.meta,positioning,calc,fit}
\usepackage{pgfplots}
\pgfplotsset{compat=1.18}
\usepackage[numbers,sort&compress]{natbib}
\usepackage{hyperref}
\usepackage[capitalise,noabbrev,nameinlink]{cleveref}

\definecolor{qblue}{HTML}{1F4E79}     
\definecolor{aiorange}{HTML}{E69F00}  
\definecolor{qorange}{HTML}{D55E00}   
\definecolor{coopt}{HTML}{009E73}     
\colorlet{aiqo}{qblue}                
\definecolor{mat1}{HTML}{DCE9F5}      
\definecolor{mat2}{HTML}{B5D0EA}
\definecolor{mat3}{HTML}{7FB0DA}
\definecolor{mat4}{HTML}{4A8BC4}
\definecolor{mat5}{HTML}{2166AC}
\definecolor{mat6}{HTML}{0B3C73}

\newcommand{\E}{\mathbb{E}}
\newcommand{\Dist}{\mathcal{D}}
\newcommand{\Ctrain}{C_{\mathrm{train}}}
\newcommand{\Cinfer}{C_{\mathrm{infer}}}
\newcommand{\Copt}{C_{\mathrm{opt}}}
\newcommand{\Ctotal}{C_{\mathrm{total}}}
\newcommand{\NQPU}{N_{\mathrm{QPU}}}
\newcommand{\Nshots}{N_{\mathrm{shots}}}
\newcommand{\aQ}{\mathsf{Q}}
\newcommand{\aC}{\mathsf{C}}
\newcommand{\AIxQ}{AI\,$\times$\,Quantum}
\newcommand{\nCore}{57}

\newcommand{\nStrongBase}{18}
\newcommand{\nLearnedBase}{11}
\newcommand{\nNoBase}{22}
\newcommand{\nQAblApplicable}{24}
\newcommand{\nQAblYes}{10}
\newcommand{\nQAblPartial}{9}
\newcommand{\nQAblNo}{5}
\newcommand{\nQubitsStated}{39}
\newcommand{\nQubitsLeTwenty}{25}
\newcommand{\nQubitsGtHundred}{4}
\newcommand{\medQubits}{16}
\newcommand{\nHardware}{14}
\newcommand{\nResCoded}{52}
\newcommand{\nShotsReported}{18}
\newcommand{\nExactExp}{12}
\newcommand{\nShotsNotReported}{21}
\newcommand{\nCallsReported}{14}

\newcommand{\nBothResources}{9}
\newcommand{\nParadigm}{52}
\newcommand{\nQforAI}{15}

\newcommand{\nQforAILearned}{9}
\newcommand{\nQforAIQAblYes}{5}
\newcommand{\nQforAIQAblApp}{15}
\newcommand{\nQforAIHardware}{2}
\newcommand{\nAIforQO}{27}
\newcommand{\nAIforQONoBase}{18}
\newcommand{\nCoopt}{10}

\newcolumntype{L}{>{\raggedright\arraybackslash}X}
\newcolumntype{P}[1]{>{\raggedright\arraybackslash}p{#1}}

\begin{document}

\title{Beyond QAOA: A Review of AI and Quantum Computing for Adaptive
  Combinatorial Optimization}

\author{Hoong Chuin Lau}
\affiliation{School of Computing and Information Systems, Singapore
  Management University, Singapore}

\begin{abstract}
Near-term quantum approaches to combinatorial optimization are constrained by
qubit counts, circuit fidelity, sampling cost, calibration drift, and the
difficulty of encoding constraints. In parallel, machine learning is
increasingly used to configure and control quantum optimization workflows.
We call such workflows \emph{adaptive}: decisions that are conventionally
fixed in advance, from formulation and penalties to shot budgets, backends,
and whether to invoke a quantum processor at all, are instead made by learned
policies that respond to the instance, the progress of the solve, or the state
of the hardware. This review examines that intersection through three paradigms: AI for
quantum optimization, quantum for AI-driven optimization, and AI--quantum
co-optimization. Rather than organizing the literature by application, we
organize it by the decision being learned, from parameter initialization,
query selection, and circuit design, through formulation, constraint handling,
decomposition, and sampling, to hardware execution, solver selection, and
solver-policy control. A structured search identified a 119-paper reference
corpus, of which 67 papers form an evidence core coded in detail. The
evidence is considerably stronger for AI-assisted quantum optimization than
for the reverse direction. Learning already reduces
quantum evaluations, improves initialization, supports decomposition and
penalty control, mitigates noise, and informs execution choices, whereas
evidence that quantum computation improves learned optimizers remains largely
confined to small-scale simulation. Coding the evidence core on fixed
categories also shows thin experimental controls: of the \nParadigm{} studies
in the three paradigms, \nNoBase{} include no classical baseline, the quantum
contribution is fully isolated in \nQAblYes{} of the \nQAblApplicable{} where
an ablation applies, and the median of the largest experiment reported per
study is \medQubits{} qubits. We
introduce an M0--M5 evidence hierarchy
that separates simulation, generalization, hardware validation, end-to-end
execution, and practical quantum advantage under matched resources, and we
find no broadly
convincing result at the highest level in the screened corpus. We argue that
scaling is increasingly a systems problem: the question is not only whether a
problem fits on a quantum processor, but how classical and quantum resources
should be allocated across the optimization process. The review is aimed at
researchers in quantum computing, machine learning, and operations research;
two short primers make it accessible to readers from either side.
\end{abstract}

\maketitle

\section{Introduction}
\label{sec:intro}

Combinatorial optimization underpins planning and allocation decisions in
logistics, transportation, supply chains, telecommunications, manufacturing,
healthcare, and public-sector operations. Problems such as the traveling
salesman problem (TSP), the capacitated vehicle routing problem (CVRP),
maximum independent set (MIS), MaxCut, the quadratic assignment problem
(QAP), and the multidimensional knapsack problem (MDKP) remain difficult
despite substantial progress in mathematical programming, approximation
algorithms, constraint programming, local search, metaheuristics, and
learning-based optimization.

Quantum computing adds a different computational substrate to this landscape.
Near-term variational gate-based methods, namely the Quantum Approximate
Optimization Algorithm (QAOA) \citep{farhi2014,hadfield2019}, the Variational
Quantum Eigensolver (VQE) \citep{peruzzo2014}, and Quantum Random Access
Optimization (QRAO) \citep{fuller2021}, together with quantum annealing
\citep{kadowaki1998,albash2018} and analog optimization, have motivated a
large literature on discrete optimization. Existing reviews already cover the
algorithmic and complexity-theoretic foundations in depth
\citep{abbas2024,blekos2024}. They also make a point that is central to this
review: problem hardness does not by itself imply useful quantum computation,
and meaningful evaluation must account for the resources required to obtain a
solution.

Our focus is narrower. We ask how AI changes the design and operation of
quantum combinatorial optimization, and where quantum computation changes an
AI-based optimizer. The question arises because practical workflows are
rarely a single quantum call. A problem may be reformulated and decomposed
classically, mapped to one or more quantum subproblems, executed under a
sampling budget, repaired by classical optimization, and reconfigured as
diagnostics become available.

In many hybrid workflows, performance depends as much on the decisions made
around the quantum processing unit (QPU) as on the quantum routine itself.
These decisions include the choice of formulation and penalties,
decomposition, initialization, circuit structure, measurement allocation,
backend selection, and whether to use a quantum routine at all. Many of them
can be posed as learning or sequential decision problems. Treating them
explicitly also changes how results should be interpreted: reducing QPU calls,
improving feasibility, or selecting a better backend is useful even when no
practical quantum advantage has been established.

We use \emph{adaptive} in this sense. An adaptive pipeline is one in which
decisions that are conventionally fixed in advance (formulation, penalties,
decomposition, circuit structure, shot budget, backend, and whether to invoke
a QPU at all) are made by learned policies that respond to the instance, the
progress of the solve, or the state of the hardware. This is broader than the
use of the term in specific algorithms such as adaptive variants of QAOA and
VQE, which grow the circuit during optimization. Most current work adapts a
single such decision; the review traces the move toward pipelines that adapt
several, without implying that fully adaptive solvers already exist.

\begin{figure}[t]
  \centering
  \resizebox{\linewidth}{!}{\input{figures/fig0_paradigms}}
  \caption{The three paradigms differ in where learning and quantum
    computation sit. In AI-for-QO, a learned model sets one decision of a
    quantum optimizer, using quantum evaluations as feedback. In Q-for-AI, a
    quantum layer is part of the learned model itself. In AI--quantum
    co-optimization, a solver policy either coordinates several coupled
    decisions or repeatedly allocates work between quantum and classical
    actions as the search state evolves (\cref{sec:coopt}).}
  \label{fig:paradigms}
\end{figure}
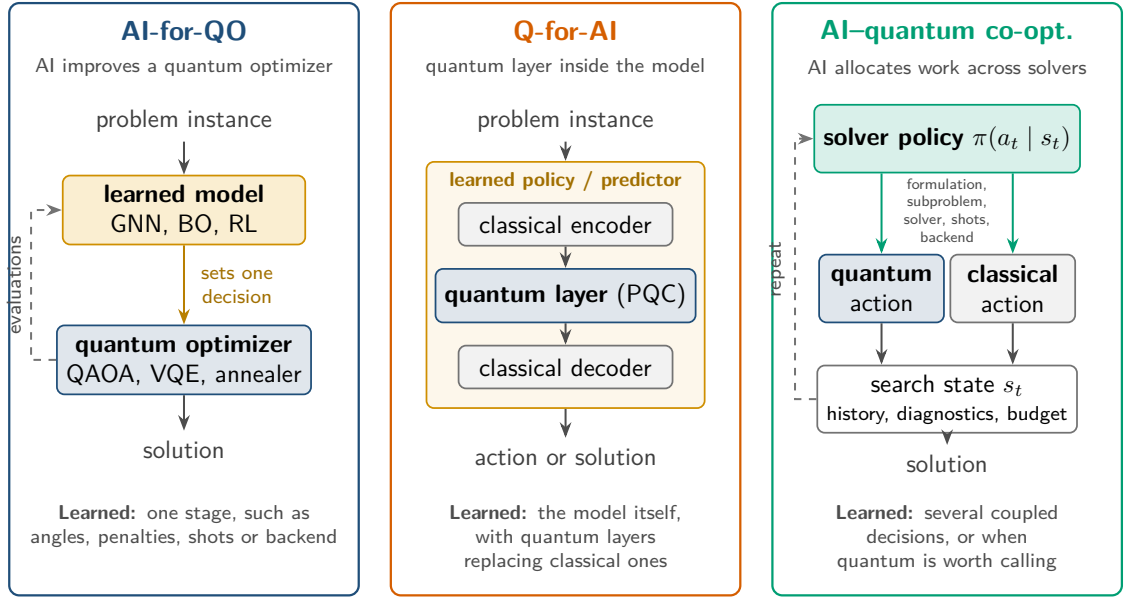

We organize the literature into three paradigms.
\emph{AI for Quantum Optimization} (AI-for-QO) uses AI to improve a component
of a quantum or hybrid quantum optimization pipeline.
\emph{Quantum for AI-Driven Optimization} (Q-for-AI) uses quantum computation
as part of a learned model or policy for combinatorial optimization.
\emph{AI--Quantum Co-Optimization} uses AI either to coordinate several
interacting decisions or to allocate work repeatedly between classical and
quantum solvers as the search evolves. The three categories
make different empirical claims. Predicting QAOA parameters concerns the
configuration of a quantum optimizer; using an equivariant quantum circuit as
a routing policy concerns the learned model itself; and choosing between
classical and quantum repair concerns resource allocation. \Cref{fig:paradigms} contrasts
where learning and quantum computation sit in each paradigm.

\begin{figure}[t]
  \centering
  \input{figures/fig1_progression}
  \caption{Increasing scope of learned decisions in quantum combinatorial
    optimization. Each stage widens what is learned (labels below each box),
    from a fixed quantum algorithm to policies that allocate work across
    heterogeneous classical and quantum resources.}
  \label{fig:progression}
\end{figure}
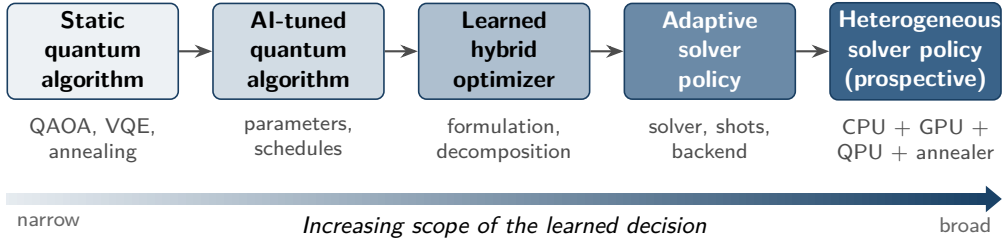

The literature can be organized by the scope of the decision being learned:
some work learns parameters for a fixed circuit; other work learns circuit
structure, formulation or decomposition choices, and execution settings; and
some systems learn conditional solver policies over classical and quantum
actions. \Cref{fig:progression} summarizes these levels of scope. The
ordering is conceptual rather than chronological, and it is not a claim that
all methods follow the same path.

The argument of the review follows from this view. If the decisions made
around the QPU matter as much as the quantum routine, the most consequential
of them is whether to call the QPU at all, a choice we refer to as
\emph{selective quantum use}. It takes two forms. \emph{Ex-ante solver
selection} decides before solving whether an instance should go to a quantum
or a classical solver; \emph{online selective invocation} decides during an
ongoing hybrid search whether a particular quantum call is worth its cost.
The first is established in the literature; the second is rare. We show that learning already improves many of
these decisions (\cref{sec:aiqo,sec:qai,sec:coopt}); that the evidence behind
each paradigm is real but shallow, and that its limits are set more by the
whole system than by qubit counts (\cref{sec:maturity}); and that selective
quantum use is both the principle shared by the most informative recent
systems and the open problem whose resolution would most strengthen the
evidence (\cref{sec:alns,sec:when-not}).

\paragraph{Contributions.} This review makes four contributions:
\begin{enumerate}
  \item It introduces a \emph{pipeline taxonomy} that classifies work by the
        optimization decision being learned (\cref{sec:taxonomy}).
  \item It separates AI-for-QO, Q-for-AI, and AI--Quantum Co-Optimization and
        compares the strength of evidence across the three
        (\cref{sec:aiqo,sec:qai,sec:coopt}).
  \item It proposes an \emph{M0--M5 evidence hierarchy} that distinguishes
        simulation, generalization, hardware validation, end-to-end
        execution, and practical quantum advantage under matched resources
        (\cref{sec:maturity}).
  \item It develops benchmarking principles and research priorities around
        selective quantum use, strong classical baselines, resource
        accounting, and cross-problem generalization
        (\cref{sec:benchmarking,sec:agenda}).
\end{enumerate}

\subsection{Summary of Main Findings}
\label{sec:summary}

For readers who want the conclusions first, the main findings are as follows.
\begin{itemize}
  \item \textbf{The evidence is asymmetric.} AI-for-QO has the largest body of
        hardware-grounded results: learned initialization, query-efficient
        parameter search, noise mitigation, annealing-schedule design, and
        hardware-aware execution. Seven of the \nAIforQO{} AI-for-QO studies coded in
        detail reach at least the M2--M3 level. Q-for-AI is less mature: 13 of
        15 coded studies remain at M1--M2. Most Q-for-AI studies
        (\nQforAILearned{} of \nQforAI{}) do compare
        against a classical learned model, but only \nQforAIQAblYes{} replace
        the quantum layer with a classical one at matched size, and only
        \nQforAIHardware{} execute on physical hardware (\cref{sec:asymmetry}).
  \item \textbf{Experimental controls are thin.} Of the \nParadigm{} studies
        in the three paradigms, \nNoBase{} include no
        classical baseline of any kind and \nStrongBase{} compare against a
        strong heuristic or exact solver. The quantum contribution is fully
        isolated in \nQAblYes{} of the \nQAblApplicable{} studies where an
        ablation applies. Experiments are small: the median of the largest
        experiment reported per study is \medQubits{} qubits, and
        \nQubitsLeTwenty{} of \nQubitsStated{} studies stay at 20 qubits or fewer;
        only \nCallsReported{} of \nResCoded{} studies report the number
        of quantum evaluations or QPU calls (\cref{sec:baselines}).
  \item \textbf{No study reaches practical quantum advantage.} The
        screened corpus contains no broadly convincing M5 result in any
        paradigm. Reported benefits are typically reductions in quantum
        resources or improvements over other quantum baselines, which are
        useful but distinct from practical quantum advantage.
  \item \textbf{Constraints, not qubit count alone, limit scaling.} Penalty
        conditioning, slack variables, sparse feasible subspaces, transpiled
        two-qubit gate counts, and shot budgets often dominate performance on
        constrained problems (\cref{sec:scaling}).
  \item \textbf{Selective quantum use is emerging as a design principle.}
        Ex-ante solver selection is established, but online selective
        invocation, in which a learned controller declines an individual
        quantum call when its expected marginal value does not justify its
        cost, is rare (\cref{sec:alns,sec:when-not}).
\end{itemize}
\noindent The main limitations are that the field is moving quickly, much of
the 2025--2026 literature consists of preprints, and reported metrics are too
heterogeneous for a quantitative meta-analysis (\cref{sec:threats}). The
maturity levels record what has been demonstrated, not the scientific quality
of individual studies.

\paragraph{Reading guide.} The review has three parts. The first sets up the
foundations: \cref{sec:background} provides two short primers, one on quantum
optimization for readers from AI and operations research and one on the
learning methods used in this literature for readers from quantum computing,
and \cref{sec:method} positions the review and describes how the literature
was collected and coded. The second part surveys the literature through the
pipeline taxonomy of \cref{sec:taxonomy}: \cref{sec:aiqo,sec:qai,sec:coopt}
cover the three paradigms, and \cref{sec:problems} cuts across them by
problem family. The third part assesses the evidence: \cref{sec:maturity}
grades what has been demonstrated and explains why it stalls,
\cref{sec:benchmarking} turns the gaps into benchmarking principles, and
\cref{sec:agenda} states the open problems and the outlook for the field.

\section{Background}
\label{sec:background}

This section introduces the concepts used in the rest of the review. The
first six subsections are written for readers without a quantum background
and emphasize what each concept means operationally (what goes in, what comes
out, and what it costs) rather than the underlying physics.
\Cref{sec:ml-primer} serves the opposite audience by summarizing the
learning methods that recur in the literature. \Cref{tab:glossary} at the end
of the section summarizes the quantum vocabulary.

\subsection{From Combinatorial Problems to QUBO and Ising Models}
\label{sec:qubo}

Many of the direct quantum optimization methods reviewed here operate on the
\emph{quadratic unconstrained binary optimization} (QUBO) problem or its
equivalent Ising formulation. The QUBO problem is
\begin{equation}
  \min_{\mathbf{x}\in\{0,1\}^n} \; C(\mathbf{x}) = \mathbf{x}^{\top} Q\, \mathbf{x}
  = \sum_{i} Q_{ii}\, x_i + \sum_{i<j} (Q_{ij}+Q_{ji})\, x_i x_j ,
  \label{eq:qubo}
\end{equation}
where $Q\in\mathbb{R}^{n\times n}$ and we use $x_i^2 = x_i$ for binary
variables. (Bold $\mathbf{x}$ denotes a binary decision vector; plain $x$ in
\cref{sec:coopt} denotes a problem instance.) The same problem can be written
in terms of spins $z_i\in\{-1,+1\}$ via the substitution $x_i = (1-z_i)/2$,
which yields an \emph{Ising model}
\begin{equation}
  C(\mathbf{z}) = \sum_{i} h_i z_i + \sum_{i<j} J_{ij} z_i z_j + \text{const}.
  \label{eq:ising}
\end{equation}
The two forms are equivalent, and Ising formulations are known for many
NP-hard problems \citep{lucas2014}. MaxCut illustrates why it is the standard
benchmark: for a graph $G=(V,E)$, the number of edges cut by the partition
encoded in $\mathbf{z}$ is $\sum_{(i,j)\in E} (1-z_i z_j)/2$, which is already
in Ising form with no constraints.

Most operational problems are constrained, and QUBO has no constraints.
Constraints are therefore typically moved into the objective as penalties. An
equality constraint $A\mathbf{x}=\mathbf{b}$ becomes
\begin{equation}
  \min_{\mathbf{x}\in\{0,1\}^n} \; f(\mathbf{x})
  + \lambda\, \lVert A\mathbf{x}-\mathbf{b}\rVert_2^2 ,
  \label{eq:penalty}
\end{equation}
for a penalty weight $\lambda>0$ that must be large enough to make violations
unattractive but not so large that it swamps the objective. An inequality
constraint such as a knapsack capacity $\mathbf{w}^{\top}\mathbf{x}\le W$
is usually converted to an equality by adding binary \emph{slack variables},
which consume additional qubits. These are exactly the formulation decisions
discussed in \cref{sec:formulation}: the penalty weight, the slack encoding,
and the resulting number of variables directly determine how hard the problem
is for the quantum device. In the standard mapping, each binary variable
occupies one qubit, so a TSP on $n$ cities with the usual position-based
encoding needs $n^2$ qubits before any further constraint handling. An
alternative to penalties is to restrict the dynamics to the feasible subspace
using constraint-preserving mixers \citep{hadfield2019}, at the cost of more
complex circuits.

\subsection{Qubits, Measurement, and Shots}
\label{sec:shots}

A register of $n$ \emph{qubits} can be prepared in a \emph{superposition} that
assigns a complex amplitude $\alpha_{\mathbf{x}}$ to each of the $2^n$
computational-basis states (bitstrings) $\mathbf{x}$. For the purposes of this
review, the essential fact is what happens at the end: \emph{measuring} the
register in the computational basis returns a single bitstring, drawn at
random with probability $|\alpha_{\mathbf{x}}|^2$. A quantum optimizer is
therefore best thought of as a programmable \emph{sampler} over candidate
solutions. Each preparation and measurement is called a \emph{shot}.
Quantities such as the expected cost
\begin{equation}
  F = \E_{\mathbf{x}\sim|\alpha|^2}\bigl[C(\mathbf{x})\bigr]
  \;\approx\; \hat F = \frac{1}{S}\sum_{s=1}^{S} C\bigl(\mathbf{x}^{(s)}\bigr)
  \label{eq:shots}
\end{equation}
must be estimated from $S$ shots, with a statistical error that shrinks only
as $O(1/\sqrt{S})$. This is why shot budgets, and the allocation of shots
across calls, appear throughout the review as first-class resources
(\cref{sec:query,sec:agentic}).

On \emph{gate-based} devices, a computation is specified as a \emph{circuit}:
a sequence of one- and two-qubit operations (\emph{gates}) applied to the
register. The \emph{depth} of a circuit is the number of sequential gate
layers. Physical devices connect only certain pairs of qubits, so a circuit is
\emph{transpiled} (compiled) onto the hardware, which typically inserts extra
two-qubit gates to move information between non-adjacent qubits. Two-qubit
gates are typically the dominant source of error on current hardware, so the
\emph{transpiled two-qubit gate count}, rather than the number of logical
variables, is often what limits the problems that can be run usefully
(\cref{sec:scaling}). Current devices are described as \emph{noisy
intermediate-scale quantum} (NISQ) hardware \citep{preskill2018,bharti2022}: they
typically offer tens to a few hundred usable qubits, lack full error
correction, and produce outputs that approach uniform noise as circuits grow
deeper.

\subsection{Quantum Annealing}
\label{sec:annealing}

Quantum annealing \citep{kadowaki1998} is the quantum counterpart of
simulated annealing. Where simulated annealing escapes local minima using
thermal fluctuations that are gradually reduced, quantum annealing uses
quantum fluctuations. The device implements a time-dependent Hamiltonian
\begin{equation}
  H(s) = \bigl(1-s\bigr)\, H_{M} + s\, H_{C}, \qquad s = t/T \in [0,1],
  \label{eq:annealing}
\end{equation}
that interpolates over an anneal time $T$ from a \emph{mixer} Hamiltonian
$H_M = -\sum_i X_i$, whose ground state is the uniform superposition over all
bitstrings ($X_i$ is the Pauli-$X$, or bit-flip, operator on qubit $i$), to a
\emph{problem} Hamiltonian $H_C$, which is diagonal in the computational basis
and whose ground state encodes the optimal solution of \cref{eq:ising}.
Under the assumptions of the adiabatic theorem, sufficiently slow evolution
keeps the system close to the instantaneous ground state and can therefore
prepare the ground state of $H_C$ \citep{albash2018}. The required anneal time grows
as the minimum spectral gap along the path shrinks (in the simplest analyses,
roughly as the inverse square of that gap), and the gap can close rapidly for
hard instances. Practical anneals are therefore fast and return a
distribution of good but not necessarily optimal bitstrings, and many runs
(reads) are taken, as with shots. Commercial annealers, such as those from
D-Wave, are analog devices with thousands of qubits but sparse connectivity,
so each logical variable is usually represented by a \emph{chain} of physical
qubits found by \emph{minor embedding}. The schedule $s(t)$, the embedding,
and the chain strength are all tunable, which makes them natural targets for
learning (\cref{sec:hardware}).

\subsection{The Quantum Approximate Optimization Algorithm}
\label{sec:qaoa-primer}

QAOA \citep{farhi2014} can be understood as a discretized, parameterized
version of annealing that runs on gate-based hardware. Starting from the
uniform superposition $|+\rangle^{\otimes n}$, it alternately applies the
problem and mixer operators $p$ times:
\begin{equation}
  |\psi(\boldsymbol{\gamma},\boldsymbol{\beta})\rangle
  = \prod_{\ell=1}^{p} e^{-i\beta_\ell H_M}\, e^{-i\gamma_\ell H_C}\;
    |+\rangle^{\otimes n},
  \qquad H_M = \sum_{i=1}^{n} X_i ,
  \label{eq:qaoa}
\end{equation}
where the product is applied with $\ell=1$ first. The mixer is the same as in
\cref{eq:annealing} up to a sign, which does not matter here because the
angles $\beta_\ell$ can take either sign. The $2p$ real angles
$\boldsymbol{\theta} = (\boldsymbol{\gamma},\boldsymbol{\beta})$ are the
algorithm's only free parameters, and $p$ is called the \emph{depth} or number
of \emph{layers}. The angles are chosen to optimize (for a minimization
problem, minimize) the expected cost
$F(\boldsymbol{\theta}) =
\langle\psi(\boldsymbol{\theta})|H_C|\psi(\boldsymbol{\theta})\rangle$, which
is the quantity estimated by shots in \cref{eq:shots}. Solution quality is
commonly reported as the \emph{approximation ratio}
$r = F(\boldsymbol{\theta})/C^{\star}$ for maximization problems such as
MaxCut, where $C^{\star}$ is the optimal value.

\begin{figure}[t]
  \centering
  \input{figures/fig_variational_loop}
  \caption{The hybrid variational loop used by QAOA and other variational
    quantum algorithms. The quantum processor acts as a parameterized sampler;
    all optimization of the parameters happens classically, using objective
    estimates obtained from $S$ shots.}
  \label{fig:variational-loop}
\end{figure}
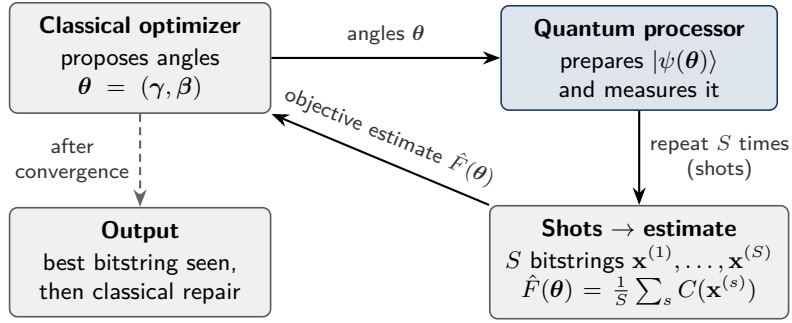

The optimization of $\boldsymbol{\theta}$ is carried out by a
\emph{classical} optimizer in a hybrid loop (\cref{fig:variational-loop}):
the optimizer proposes angles, the quantum processor prepares the state and
returns $S$ samples, the sample mean is fed back, and the process repeats.
From a classical perspective, QAOA is therefore a randomized heuristic with
$2p$ tunable parameters whose objective can be evaluated only through noisy,
costly sampling. This viewpoint explains why so much of the AI-for-QO
literature (\cref{sec:params,sec:query}) focuses on predicting good angles or
reducing the number of evaluations: each loop iteration consumes QPU time and
shots. With a suitable sequence of parameters, increasing $p$ allows QAOA to
approximate a discretized adiabatic evolution \citep{farhi2014}; this does not
imply that those parameters can be found efficiently, and larger $p$ means
deeper circuits and more noise. At low depth, QAOA performance is also subject to
structural limitations \citep{bravyi2020}, and good angles often follow
regular patterns across instances \citep{zhou2020}, which is the empirical
basis for parameter transfer.

\paragraph{Variants used in this review.} Several variants recur in later
sections. The \emph{Variational Quantum Eigensolver} (VQE)
\citep{peruzzo2014} uses the same loop with a different, often
hardware-efficient, circuit template (an \emph{ansatz}) instead of the
alternating structure of \cref{eq:qaoa}. \emph{CVaR} objectives
\citep{barkoutsos2020} replace the sample mean in \cref{eq:shots} with the
mean of the best $\alpha$-fraction of samples, which better reflects the goal
of finding one good solution rather than a good average. \emph{Recursive
QAOA} \citep{bravyi2020} uses QAOA output to fix or correlate variables one
at a time, shrinking the problem. \emph{Warm-started} QAOA initializes the
state or mixer from a classical relaxation \citep{egger2021}. \emph{Quantum
Random Access Optimization} (QRAO) \citep{fuller2021} encodes several binary
variables into each qubit, reducing the qubit count at the price of a more
complex decoding (rounding) step. More generally, these methods belong to the
family of \emph{variational quantum algorithms} (VQAs): parameterized circuits
trained by classical optimizers \citep{cerezo2021}.

\subsection{Parameterized Quantum Circuits as Learning Models}
\label{sec:pqc}

The same parameterized circuits can be used as machine-learning models rather
than as optimizers. A \emph{parameterized quantum circuit} (PQC), sometimes
called a \emph{quantum neural network} (QNN), encodes an input (for example,
a graph) into the circuit, applies trainable gates, and reads out measured
expectation values as features, scores, or action probabilities. In the
Q-for-AI work of \cref{sec:qai}, such circuits replace components of a
classical policy network, for example the attention layer of a routing model.
\emph{Equivariant} circuits are designed so that relabeling the nodes of the
input graph relabels the output in the same way, mirroring the symmetry built
into graph neural networks.

\subsection{Trainability, Noise, and Quantum Advantage}
\label{sec:advantage-primer}

Two considerations shape how results should be read. First, the training
landscapes of many parameterized circuits exhibit \emph{barren plateaus}:
gradients concentrate exponentially in the number of qubits
\citep{mcclean2018}, which can make optimization prohibitively difficult
unless the circuit structure, initialization, or training strategy avoids the
corresponding concentration regime (\cref{sec:trainability}). Hardware noise can induce a similar flattening
\citep{wang2021}. Second, we distinguish three terms that are often
conflated. \emph{Hardware execution} means that part of a workflow runs on a
physical quantum processor. \emph{Practical quantum advantage} means that a
quantum or hybrid method outperforms the strongest available classical
alternatives on a well-defined task under matched resources, in the regime
tested. \emph{Asymptotic quantum advantage} is a complexity-theoretic
statement about how resource requirements scale with problem size. Neither
hardware execution nor outperforming another quantum method establishes
practical advantage, and practical advantage in a tested regime does not
establish asymptotic advantage. The highest level of the evidence hierarchy,
M5 (\cref{sec:maturity}), corresponds to practical quantum advantage, which
is why the review insists on matched classical baselines
(\cref{sec:baselines}).

\subsection{Learning Methods for the Quantum Reader}
\label{sec:ml-primer}

The AI methods in this literature fall into a small number of families. Each
is characterized here by what it takes as input and what it outputs.
\begin{itemize}
  \item \textbf{Supervised learning and graph neural networks (GNNs).} A model
        is trained on solved examples to map an input to a prediction. GNNs
        operate on graphs by repeatedly aggregating information from each
        node's neighbours; because the aggregation does not depend on node
        labels, the predictions are permutation-equivariant. In this
        literature, GNNs map a problem graph to predicted QAOA angles, to a
        coarsened graph, or to features used for solver selection.
  \item \textbf{Bayesian optimization (BO) and surrogate models.} A
        probabilistic model, often a Gaussian process, approximates an
        expensive objective from a few evaluations, and an acquisition
        function chooses the next point to evaluate. BO is designed for
        settings with tens to hundreds of evaluations, which matches the
        cost of querying a QPU.
  \item \textbf{Reinforcement learning (RL).} An agent interacts with an
        environment modelled as a Markov decision process with states,
        actions, and rewards, and learns a policy $\pi(a\mid s)$ that
        maximizes expected cumulative reward. Deep Q-networks (DQNs) learn
        the value of each action; policy-gradient methods learn $\pi$
        directly. RL is used for sequential decisions such as circuit
        construction, penalty updates, shot allocation, and choosing between
        classical and quantum repair.
  \item \textbf{Bandits.} A single-step special case of RL in which the
        learner repeatedly chooses among options (for example, backends or
        configurations) and balances exploring uncertain options against
        exploiting good ones.
  \item \textbf{Meta-learning and transfer learning.} Models are trained
        across many instances so that knowledge carries over to new instances
        with little or no additional optimization. This is the basis of
        amortized parameter prediction (\cref{sec:params}).
  \item \textbf{Neural combinatorial optimization (NCO).} Learned policies,
        typically attention-based networks trained by RL, construct solutions
        to problems such as the TSP or vehicle routing step by step
        \citep{vinyals2015,khalil2017,kool2019}; Bengio et
        al.~\citep{bengio2021} review the field. These are the natural
        classical baselines for the Q-for-AI work of \cref{sec:qai}.
  \item \textbf{Large language model (LLM) agents.} Language models that
        propose code or configuration changes, which are then evaluated by an
        external harness. In this literature they are used to search over
        circuit designs and solver policies (\cref{sec:agentic}).
\end{itemize}

\begin{table}[t]
  \caption{Glossary of quantum-computing terms used in this review, with the
    closest classical analogy.}
  \label{tab:glossary}
  \small
  \begin{tabularx}{\linewidth}{@{}P{2.7cm}LP{3.8cm}@{}}
    \toprule
    \textbf{Term} & \textbf{Meaning in this review}
      & \textbf{Classical analogy} \\
    \midrule
    Qubit & Quantum bit; one per binary decision variable in a standard QUBO
      mapping & Binary variable \\
    QUBO / Ising & Unconstrained quadratic objective over binary or $\pm1$
      variables (\cref{eq:qubo,eq:ising}) & Quadratic 0--1 program \\
    Hamiltonian & Operator whose energy encodes an objective; $H_C$ encodes
      the cost & Objective function \\
    Shot & One preparation and measurement, returning one bitstring
      & One random sample \\
    Circuit depth & Number of sequential gate layers & Sequential steps \\
    Two-qubit gate & Interaction between two qubits; main error source
      & Costly, error-prone operation \\
    Transpilation & Compiling a circuit to a device's connectivity and gate
      set & Compilation to target hardware \\
    QPU & Quantum processing unit, usually accessed remotely
      & Accelerator (GPU) \\
    NISQ & Noisy intermediate-scale quantum hardware without full error
      correction & Unreliable, small accelerator \\
    Annealing schedule & Time profile $s(t)$ of the interpolation in
      \cref{eq:annealing} & Cooling schedule \\
    Minor embedding & Mapping each logical variable to a chain of physical
      qubits & Graph embedding \\
    Ansatz & Fixed circuit template with trainable parameters
      & Model architecture \\
    QAOA depth $p$ & Number of alternating problem/mixer layers
      & Number of heuristic iterations \\
    Barren plateau & Exponentially vanishing gradients in large circuits
      & Vanishing gradients \\
    Practical quantum advantage & Outperforming the strongest classical
      alternatives under matched resources, in a tested regime
      & State of the art under matched budget \\
    \bottomrule
  \end{tabularx}
\end{table}

\section{Scope and Review Methodology}
\label{sec:method}

This section positions the review against existing reviews, describes how
the literature was collected, screened, and coded, and states the research
questions that organize the rest of the paper.

\subsection{Relationship to Existing Reviews}
\label{sec:scope}

Abbas et al.~\citep{abbas2024} provide the broadest recent foundation for quantum
optimization, spanning complexity, exact and approximate methods, heuristics,
mixed-integer optimization, hardware, and benchmarking. Blekos et al.~\citep{blekos2024}
focus more specifically on QAOA and its variants. Gemeinhardt et al.~\citep{gemeinhardt2023}
systematically map quantum combinatorial optimization in the NISQ era, while
Yarkoni et al.~\citep{yarkoni2022} review quantum annealing for industrial applications.

A second adjacent literature concerns AI for quantum computing in general.
Alexeev et al.~\citep{alexeev2025} survey AI across the quantum stack, including control,
circuit synthesis, characterization, error correction, and post-processing.
Quantum architecture search is mature enough to have dedicated reviews
\citep{martyniuk2024}, and variational quantum algorithms and their
trainability are covered comprehensively by Cerezo et al.~\citep{cerezo2021}.
Bharti et al.~\citep{bharti2022} review near-term algorithms more broadly,
including variational, annealing-inspired, and quantum machine-learning
methods and the error-mitigation and benchmarking practices that accompany
them.

The present review is deliberately narrower in application domain but broader
in workflow. Its unit of analysis is the complete combinatorial-optimization
process: what is learned, where the quantum component enters, what resources
are consumed, and what the experiments actually establish.
\Cref{tab:positioning} positions this scope against the closest reviews. The
aim is not to repeat QAOA taxonomies or generic AI-for-quantum surveys, but to
examine learning across the optimization pipeline and to grade the evidence
for each claim.

\begin{table}[t]
  \caption{Positioning relative to the closest review literature.}
  \label{tab:positioning}
  \small
  \begin{tabularx}{\linewidth}{@{}P{2.6cm}LLP{1.35cm}L@{}}
    \toprule
    \textbf{Review} & \textbf{Primary scope} & \textbf{AI interventions}
      & \textbf{Q-for-AI} & \textbf{Workflow / evidence synthesis} \\
    \midrule
    Abbas et al.~\citep{abbas2024} & Quantum optimization broadly & Limited / distributed
      & No & Complexity, algorithms, benchmarking \\
    Blekos et al.~\citep{blekos2024} & QAOA and variants & Parameter / variant coverage
      & No & Algorithm-centric \\
    Gemeinhardt et al.~\citep{gemeinhardt2023} & NISQ combinatorial optimization & Incidental
      & No & Systematic mapping by problems and methods \\
    Alexeev et al.~\citep{alexeev2025} & AI across the quantum stack & Broad AI-for-quantum
      & Limited & Quantum-stack-centric \\
    Martyniuk et al.~\citep{martyniuk2024} & Quantum architecture search & Circuit architecture
      & No & Search-space / strategy taxonomy \\
    \midrule
    \textbf{This review} & \AIxQ{} for combinatorial optimization
      & Parameters through solver policies & Yes
      & Pipeline taxonomy + M0--M5 evidence hierarchy \\
    \bottomrule
  \end{tabularx}
\end{table}

\subsection{Search, Screening, and Coding}
\label{sec:search}

We followed a structured literature-review protocol with a cut-off date of
6~October~2026. Searches were run with Undermind (\url{https://www.undermind.ai}),
an AI-assisted semantic literature-search tool. Three broad searches targeted
(i)~AI for quantum optimization, (ii)~quantum for AI-driven optimization, and (iii)~surveys and
frontier taxonomies. They returned 179, 61, and 207 candidate records,
respectively, before deduplication; the sets overlap, so the counts are not
additive. We then ran problem-centered searches for routing, MIS/MaxCut,
knapsack/MDKP, QAP/assignment, and scheduling. After deduplication and
eligibility screening, 119 papers were retained in the reference corpus, of
which 67 form an evidence core that was coded in detail and used for the
comparative synthesis. The coding scheme is given in \cref{app:fields}.

Candidate papers were first screened by title and abstract, and central claims
were checked against the full text where available. Earlier studies were
retained when they established a method or category used by later work; the
main emphasis is on 2022--2026. Because terminology is inconsistent across AI,
quantum computing, and operations research (OR), semantic retrieval was
preferred to a narrow keyword-only query.

Coding used two tiers (\cref{app:fields}). The first tier records
descriptive fields for every core study: problem class, AI and quantum
methods, pipeline intervention, learned object, training and test regimes,
problem size, claimed benefit, the strongest conclusion supported by the
experiments, and an M0--M5 maturity level (\cref{sec:maturity}) based on
demonstrated evidence rather than perceived scientific quality. The second
tier codes four evidence fields with fixed categories, so that the counts
reported in this review can be reproduced: (F1)~the strongest classical
non-learned baseline used as a competitor (none, simple, strong, or exact);
(F2)~whether a classical learned baseline is included; (F3)~whether the
contribution of the quantum component is isolated by a classical replacement
(yes, partial, no, or not applicable when the quantum routine is itself the
object of study); and (F4)~scale and resources: maximum qubits overall and on
hardware, hardware type, and whether shots and quantum evaluations are
reported. Second-tier fields distinguish \emph{not reported} (the paper was
read and the information is absent) from \emph{not checked} (the full text
was unavailable). All \nCore{} non-survey core studies were coded from the
full text. One study first selected for the evidence core, a quantum
Q-learning model for the CVRP published in a 2024 IEEE symposium, was
moved to the reference corpus because its full text could not be obtained;
the core counts above exclude it. Counts reported in the text refer to the
\nParadigm{} studies assigned to the three paradigms; the four structural
precursors and the benchmark study are coded in the same way, and counts that
include them are given in the supplementary spreadsheet. The complete
119-paper corpus is also listed there; the bibliography contains only the
papers cited explicitly in the text.

\paragraph{Inclusion criteria.} A study was included in the analytical core if
(a)~AI materially changed the configuration or operation of a quantum
optimizer; (b)~quantum computation materially changed a learned optimizer or
policy; (c)~AI coordinated multiple classical and quantum decisions; or
(d)~the study provided essential review or methodological background.

\paragraph{Exclusion criteria.} We excluded ordinary QAOA, VQE, or annealing
studies without substantive learning; generic quantum machine learning (QML)
papers without combinatorial-optimization content; quantum-inspired classical
methods unless needed for comparison; and studies in which ``learning'' merely
denoted the use of a conventional numerical optimizer inside a standard VQA
loop.

\subsection{Research Questions}
\label{sec:rqs}

The review addresses seven research questions:
\begin{description}
  \item[RQ1] Where does AI intervene in the quantum combinatorial-optimization
    pipeline?
  \item[RQ2] What does the AI component learn: parameters, representations,
    circuits, formulations, decompositions, resources, or policies?
  \item[RQ3] What benefit is demonstrated: quality, feasibility,
    generalization, QPU calls, shots, qubits, depth, runtime, robustness, or
    cost?
  \item[RQ4] What transfers across instances, sizes, graph families, problem
    classes, noise regimes, and hardware?
  \item[RQ5] Where does quantum computation materially improve AI-based
    combinatorial optimization?
  \item[RQ6] How strong is the evidence, given classical baselines, matched
    budgets, noise, shots, hardware, held-out testing, and ablations?
  \item[RQ7] Is the literature moving from component-level learning toward
    solver-level adaptive control?
\end{description}
RQ1 and RQ2 are answered by the pipeline taxonomy (\cref{sec:taxonomy}) and
the survey of the three paradigms (\cref{sec:aiqo,sec:qai,sec:coopt}); RQ3
and RQ4 by those sections together with the problem-centered synthesis
(\cref{sec:problems}); RQ5 and RQ6 by the evidence assessment
(\cref{sec:maturity}); and RQ7 by the co-optimization section and the
outlook (\cref{sec:coopt,sec:outlook}).

\subsection{Threats to Validity and Review Limitations}
\label{sec:threats}

This review is broad but is not a formal meta-analysis. The literature is
moving quickly, especially in 2026; preprints can change before archival
publication, and the reported metrics are too heterogeneous for pooled effect
estimates. Semantic search may also under-represent adjacent work that uses
different terminology. We therefore separate the broad reference corpus from
the evidence core, use explicit inclusion boundaries, and verify central
claims against full text where possible. Maturity levels and the evidence
fields were assigned by a single coder following the rules in
\cref{app:fields}. First-pass values for the four evidence fields were
drafted with the assistance of an AI tool (Claude, Anthropic) from the open
full texts; every value was then checked by the author against the paper and
corrected where necessary. The author takes full responsibility for the coding
and the conclusions. Borderline
maturity cases are recorded as ranges rather than forced into one level. Absence of reported
evidence is not treated as evidence of failure, and maturity labels indicate
only what has been demonstrated.

\section{A Pipeline Taxonomy for \AIxQ{} Combinatorial Optimization}
\label{sec:taxonomy}

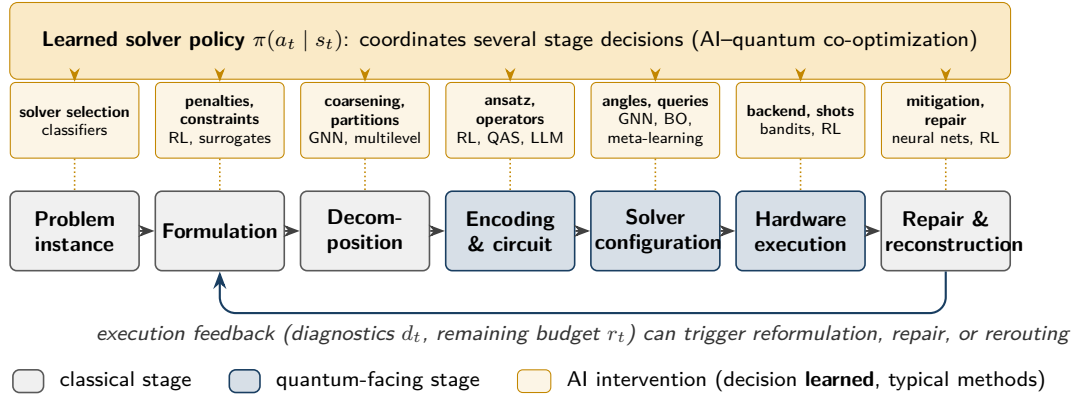
\begin{figure}[t]
  \centering
  \input{figures/fig2_pipeline}
  \caption{Pipeline view of AI intervention points and execution feedback.
    Each stage carries the decision that is learned and the AI methods
    typically used (\cref{tab:taxonomy}). Component-level methods
    (AI-for-QO) act on a single stage; co-optimization methods either learn
    several coupled decisions across stages or repeatedly allocate work
    between heterogeneous solver actions as the solver state, including
    execution feedback, evolves (\cref{sec:boundary}).}
  \label{fig:pipeline}
\end{figure}

We represent a heterogeneous optimization workflow as a sequence that runs
from the problem instance through formulation, decomposition, encoding, solver
configuration, and hardware execution to repair and reconstruction. AI can
intervene at every stage, and execution feedback can close the loop
(\cref{fig:pipeline}). This pipeline view is more discriminating than an
application taxonomy because two papers solving the same MaxCut instance may
make fundamentally different claims: one predicts parameters, another designs
circuits, and a third decides whether quantum execution should occur at all.
\Cref{tab:taxonomy} lists, for each stage, the decision learned, the typical
AI methods, the quantum component affected, and the primary metric.

\begin{table}[t]
  \caption{Pipeline taxonomy: the decision learned at each stage, typical AI
    methods, the quantum component affected, and the primary evaluation
    metric.}
  \label{tab:taxonomy}
  \small
  \begin{tabularx}{\linewidth}{@{}P{2.4cm}LLP{2.4cm}L@{}}
    \toprule
    \textbf{Pipeline stage} & \textbf{Decision learned}
      & \textbf{Typical AI methods} & \textbf{Quantum component}
      & \textbf{Primary metric} \\
    \midrule
    Parameter initialization & QAOA/VQA angles
      & GNNs, meta-learning, clustering, diffusion & QAOA/VQA
      & calls, convergence, quality \\
    Query control & next parameter point
      & Bayesian optimization, online surrogates & QAOA/VQA
      & evaluations, shots \\
    Circuit / ansatz design & operators, gates, depth, topology
      & RL, differentiable NAS, transformers/LLMs & QAOA/VQA
      & depth, calls, quality \\
    Formulation / penalties & QUBO coefficients, constraints
      & RL, learned surrogates, GNNs & QUBO for VQE, QAOA, annealing
      & feasibility, qubits, gap \\
    Reduction / decomposition & subgraphs, variables, subproblems
      & GNNs, multilevel learning, RL & local quantum subproblems
      & scale, gap, width \\
    Sampling / measurement & shots, CVaR level, recursion budget
      & RL, adaptive control & measurement loop
      & shots, confidence \\
    Error mitigation & corrected samples/estimates
      & neural models, regressors & noisy QPU output
      & TVD, sample quality \\
    Hardware execution & backend, embedding, layout
      & classifiers, bandits, RL & QPU/annealer
      & fidelity, runtime \\
    Solver selection & classical vs.\ quantum; solver family
      & supervised ML, algorithm selection & solver portfolio
      & selection accuracy, total cost \\
    Solver-policy control & conditional multi-stage action
      & RL, bandits, LLM agents & complete hybrid pipeline
      & end-to-end utility \\
    \bottomrule
  \end{tabularx}
  \par\smallskip
  {\footnotesize\raggedright Abbreviations: GNN, graph neural network;
    RL, reinforcement learning; NAS, neural architecture search; LLM, large
    language model; QUBO, quadratic unconstrained binary optimization; CVaR,
    conditional value-at-risk; TVD, total variation distance.\par}
\end{table}

The taxonomy organizes the next four sections. \Cref{sec:aiqo} covers
methods that learn a single stage of \cref{tab:taxonomy} while the others
stay fixed. \Cref{sec:qai} covers the reverse direction, in which the quantum
circuit is part of the learned model rather than a stage that the model
configures. \Cref{sec:coopt} covers policies that span several stages, the
last row of the table, or that repeatedly allocate work between classical and
quantum solvers during the search. \Cref{sec:problems} then reorganizes the same
literature by problem family.

\section{AI for Quantum Optimization}
\label{sec:aiqo}

AI-for-QO methods learn one stage of the pipeline in \cref{tab:taxonomy}
while holding the others fixed. Most studies in this section use learning
directly; \cref{sec:formulation} also discusses structural precursors that are
not themselves AI-driven but expose formulation choices that later learned
controllers can adapt. The subsections are organized by intervention point:
where a co-optimization system contains a component relevant to that
intervention, we discuss it here but retain its primary classification in
\cref{sec:coopt}. We follow the order of the pipeline: first
parameters and queries, then circuit structure, formulation, and
decomposition, then execution-level decisions, and finally the trainability
limits that bound all of them. The usual comparator in this literature is the
unassisted quantum routine, such as random initialization or a default
schedule, which shapes what the evidence can establish
(\cref{sec:asymmetry}).

\subsection{Parameter Prediction, Transfer, and Amortized Optimization}
\label{sec:params}

Parameter optimization is one of the earliest and most mature AI-for-QO
targets. It rests on a structural property of QAOA: for fixed parameters,
the objective concentrates over typical instances drawn from the same
distribution \citep{brandao2018}, and good angles follow regular patterns
across related instances \citep{zhou2020}, so parameters optimized for one
instance often transfer to others \citep{shaydulin2023}. A learned predictor
exploits the same concentration but conditions it on instance features, which
explains both why prediction works and why it degrades when test instances
leave the training distribution.
Khairy et al.~\citep{khairy2020} formulate QAOA parameter optimization as a learning
problem, using reinforcement learning and generative modeling so that
knowledge from training instances can be reused on unseen instances.
Jain et al.~\citep{jain2022} use graph neural networks to warm-start QAOA for MaxCut and
report transfer across instances and to larger graphs. Moussa et al.~\citep{moussa2022} use
unsupervised clustering and graph representations to identify good parameters
without repeatedly optimizing them from scratch, and Falla et al.~\citep{falla2024} learn
graph embeddings to identify donor instances for parameter transfer. More
recent work extends this direction through transfer learning
\citep{montanez2025}, conditional diffusion models \citep{meng2025}, and
graph-conditioned initialization models such as QSeer \citep{jiang2025}.

These methods mainly change the \emph{cost structure} of variational
optimization. Instead of re-optimizing a circuit from scratch for every
instance, part of the search cost is paid once during training and reused
later. For $N$ instances, the total cost of a learned approach is
\begin{equation}
  \Ctotal(N) = \Ctrain + N\,\Cinfer,
  \label{eq:amortization}
\end{equation}
where $\Ctrain$ is the one-off cost of training the predictor (including any
quantum evaluations used to generate training targets) and $\Cinfer$ is the
per-instance cost of prediction plus any residual fine-tuning. Compared with a
per-instance optimizer of cost $\Copt$, learning pays off only once
\begin{equation}
  N > N^{\star} = \frac{\Ctrain}{\Copt - \Cinfer},
  \qquad \Copt > \Cinfer .
  \label{eq:breakeven}
\end{equation}
Approximation quality should therefore be read together with the number of
circuit evaluations and the cost of learning the predictor; the break-even
number of instances $N^{\star}$ is a natural quantity to report
(\cref{sec:amortize}).

\subsection{Query-Efficient Quantum Optimization}
\label{sec:query}

Cheng et al.~\citep{cheng2024} introduce double adaptive-region Bayesian optimization for
QAOA and demonstrate faster, more stable parameter search than standard
optimizers, including a proof of concept on superconducting hardware.
O'Leary et al.~\citep{oleary2025} push the idea further with online surrogate learning that
requires no upfront surrogate training. They demonstrate query-efficient
optimization on a 127-qubit Ising problem and transfer the learned parameters
to still larger systems.

The main measurable benefit in these studies is a reduction in quantum
evaluations. A learned surrogate can be useful even if the final solver has no
practical quantum advantage, because QPU evaluations are scarce and costly.
For that reason, the number of hardware calls and shots should be treated as
primary experimental outcomes rather than implementation details.

\subsection{Circuit and Ansatz Design}
\label{sec:circuits}

Quantum architecture search (QAS) generalizes AI assistance from continuous
parameter tuning to discrete circuit design. Differentiable quantum
architecture search \citep{zhang2022}, QuantumDARTS \citep{wu2023}, RL-based
circuit design \citep{fodera2024}, RL-assisted recursive QAOA
\citep{patel2023}, and later architecture-search systems demonstrate that gate
sequences, operator pools, depth, and connectivity can themselves be learned.
QAOA-GPT \citep{tyagin2025} trains a transformer to generate circuit
structures and parameters conditioned on problem representations.

These methods can reduce search cost, but that result should be separated from
the stronger claim of discovering a new quantum algorithm. A model that
reproduces or assembles effective circuit structures accelerates search
without necessarily discovering anything new. Circuit imitation, search
acceleration, and algorithm discovery are different claims and should be
evaluated separately.

\subsection{Formulation, Constraints, and Penalties}
\label{sec:formulation}

Constrained combinatorial optimization exposes weaknesses that are largely
hidden by unconstrained MaxCut benchmarks. Penalty coefficients can overwhelm
the objective, auxiliary slack variables can inflate the qubit requirement,
and feasible solutions may form an exponentially small fraction of the
computational basis. Ayanzadeh et al.~\citep{ayanzadeh2020} demonstrate adaptive penalty
adjustment on a D-Wave system. More recent work applies learning to QUBO
penalty calibration, graph reduction, and constraint-aware preprocessing
\citep{rezk2026}.

A related line of work, comprising constraint-preserving mixers
\citep{hadfield2019}, feasibility-preserving encodings \citep{xie2024},
qubit-efficient relaxations \citep{fuller2021,sharma2024qrao}, and slack-free
penalties \citep{leelau2025slack,leelau2026cvar}, is not always AI-driven,
but it identifies the decisions that a learned controller may eventually need
to make. The underlying trade-offs are between qubit count, feasible-state
density, coefficient ranges, circuit depth, and sampling difficulty. These
structural choices matter before any parameter-learning method is applied.

\subsection{Reduction and Decomposition}
\label{sec:decomposition}

Available quantum hardware is far smaller than many industrial optimization
models, so multilevel and decomposition methods are used to create bounded
quantum subproblems. MLQAOA \citep{bach2024} uses graph learning to accelerate
a multilevel hybrid solver, while Rezk and Gora~\citep{rezk2026} combine GNN-guided graph
coarsening with adaptive penalties for the CVRP with time windows (CVRPTW) on
a quantum annealer. Sharma and Lau~\citep{sharma2026cvrp} use Lagrangian decomposition to
create bounded knapsack subproblems within the CVRP and couple this with
learned multiplier control and hardware-aware execution decisions.

Once decomposition is introduced, scalability depends on both the size of
the quantum subproblems and the classical coordination required between
them. The question is no longer simply whether the full model fits on a QPU,
but which part of the model should be sent to the QPU, how large that
subproblem should be, and when the coordination overhead outweighs any
benefit. These choices are instance- and hardware-dependent.

\subsection{Noise Mitigation, Hardware Selection, and Solver Selection}
\label{sec:hardware}

Sack and Egger~\citep{sack2024} use machine-learning-based error mitigation to execute QAOA
on nonplanar graphs of up to 40 qubits, demonstrating that learned
post-processing can recover a meaningful optimization signal from deep
compiled circuits. Related hardware-aware work includes Bayesian design of
annealing schedules \citep{finzgar2024}, transferable annealing protocols on
neutral-atom processors \citep{leclerc2025}, and reinforcement-learning
approaches to minor embedding for quantum annealing \citep{nembrini2026}.
Hardware selection is the next logical layer: if several processors,
embeddings, or backends are available, their relative value depends on the
subproblem, circuit, calibration state, queue, and latency.

Moussa et al.~\citep{moussa2020} formulate the explicit ``to quantum or not to quantum''
problem as algorithm selection. Volpe et al.~\citep{volpe2024} learn to select among
quantum solvers and settings over hundreds of QUBO instances, while
Thelen and Mauerer~\citep{thelen2025} predict algorithm trade-offs under non-functional
requirements. These studies mark the point at which AI begins to allocate
computing resources rather than only tune a quantum component; because each
learns a single decision, the solver choice, we classify them as AI-for-QO and
treat them as precursors of co-optimization (\cref{sec:selection}).

\subsection{Trainability Limits: Barren Plateaus and Noise}
\label{sec:trainability}

Learning does not remove the trainability limits of variational quantum
circuits. Barren plateaus can make gradients exponentially small in relevant
circuit regimes \citep{mcclean2018,cerezo2021}, and hardware noise can itself
induce concentration of the objective landscape \citep{wang2021}.
Ragone et al.~\citep{ragone2024} relate barren-plateau behavior to the dimension and
structure of the circuit's dynamical Lie algebra. These results matter for
AI-for-QO because parameter transfer, architecture search, and
query-efficient optimization all act on a landscape whose trainability may
already be structurally limited.

The practical implication is modest but important. AI can avoid poor
initializations, restrict the search space, or reduce the number of
evaluations, but such improvements should not be interpreted as evidence that
barren plateaus have been eliminated. Trainability remains a property to
measure and report.

The main literature reviewed in this section uses learning to improve a
quantum optimizer; the structural methods in \cref{sec:formulation} identify
further decisions that could be learned. The next section reverses the
direction and asks whether quantum computation
can improve a learned optimizer.

\section{Quantum for AI-Driven Combinatorial Optimization}
\label{sec:qai}

We use a stricter definition for the reverse direction. A method counts as
Q-for-AI only when quantum computation changes the learned model, policy,
representation, or inference process used by an AI method for combinatorial
optimization. Ordinary QAOA or VQE that directly optimizes a QUBO is not
Q-for-AI.

\subsection{Quantum Neural Combinatorial Optimization and Equivariant Policies}
\label{sec:equivariant}

Building on parametrized quantum policies for reinforcement learning
\citep{jerbi2021}, Sanches et al.~\citep{sanches2022} replace selected attention components
in a reinforcement-learning routing model with short quantum circuits and
show competitive performance in small VRP settings. Skolik et al.~\citep{skolik2023}
introduce permutation-equivariant quantum circuits for learning on weighted
graphs and apply them to neural combinatorial optimization for the TSP.
Subsequent routing studies explore quantum graph-attention policies
\citep{le2025} and hybrid quantum--classical neural routing models
\citep{ritt2026}. These architectures are of interest because symmetry and
parameter sharing can reduce the number of trainable parameters and, in some
settings, make it less dependent on problem size.

Any benefit in this setting must arise from the representation or inductive
bias of the quantum model, since the quantum circuit is part of the learned
policy rather than a direct optimizer. Current results support compact
parameterizations and symmetry-aware architectures in selected settings, but
they do not yet show consistent end-to-end routing gains over strong
classical neural combinatorial optimizers \citep{kool2019,bengio2021}.

\subsection{Hamiltonian-Based Quantum Reinforcement Learning}
\label{sec:qrl}

Kruse et al.~\citep{kruse2024} construct quantum reinforcement learning (QRL)
architectures from problem Hamiltonians and test them on MaxCut, unit
commitment, and knapsack. The models can train better than selected quantum
baselines, but the experiments are simulation-based and do not establish an
advantage over modern classical learned optimization. Outperforming another
quantum ansatz supports a claim about model design, but not a claim of
practical quantum advantage.

\subsection{Quantum Attention, QAP Learning, and Contextual Optimization}
\label{sec:contextual}

Correll et al.~\citep{correll2022} embed quantum neural layers in an attention-based
logistics model and execute small inference components on hardware. The
study is valuable because it demonstrates a genuine architectural
substitution, but the cost of quantum execution is high and the industrial
problem must be aggressively decomposed. Ye et al.~\citep{ye2023} formulate
QAP-related constrained optimization as a quantum learning problem, moving
from per-instance optimization toward amortized prediction across instances.

Lee and Kwon~\citep{leekwon2026} make this distinction especially explicit for contextual
combinatorial optimization. Their end-to-end quantum learning model embeds
contextual information into a QAOA-style phase separator and jointly trains
the classical context encoder and the quantum parameters. Across small
MaxCut, QAP, and matching problems, the method uses far fewer trainable
parameters than selected classical decision-focused baselines. Yet the
experiments are limited to 16- and 25-variable simulations, with no
physical-QPU evaluation. The evidence therefore supports parameter efficiency
and architectural feasibility, not practical quantum advantage.

\subsection{Model Generalization Versus System Generalization}
\label{sec:generalization}

Model-level transfer can degrade once finite-shot sampling, transpilation,
and device noise are introduced: an equivariant quantum policy may transfer
across graph sizes in ideal simulation yet lose that ability on hardware.
Sharma and Lau~\citep{sharma2025diag} diagnose this gap for cross-size transfer in quantum
reinforcement learning for the TSP, while work on quantum generative models
asks whether unseen valid or high-quality samples can still be generated
\citep{gili2023,gili2024} and whether such models can enhance combinatorial
search \citep{alcazar2024}. We therefore distinguish six kinds of transfer:
across instances, sizes, topologies, problem families, noise regimes, and
hardware. The distinction returns when the evidence is graded
(\cref{sec:asymmetry}), because most Q-for-AI results establish the first
kinds of transfer but not the last two.

Both AI-for-QO and Q-for-AI change one component of the optimizer: the
configuration of a quantum routine, or the model inside a learned policy.
The next section turns to systems in which learning coordinates several
components at once.

\section{AI--Quantum Co-Optimization: From Components to Solver Policies}
\label{sec:coopt}

Some recent systems do not fit a one-directional AI-for-QO or Q-for-AI
classification, because learning either coordinates several coupled decision
dimensions or repeatedly allocates work between heterogeneous classical and
quantum solver actions as the search state evolves. We refer to these
systems as \emph{AI--Quantum Co-Optimization}; \cref{sec:boundary} applies
the two criteria to borderline cases. A convenient abstraction describes the solver as a
sequential decision process with state $s_t$, action $a_t$, and a
system-level objective. At decision step $t$, the state is
\begin{equation}
  s_t = \bigl(x,\; h_t,\; d_t,\; r_t\bigr),
  \label{eq:state}
\end{equation}
where $x$ is the problem instance, $h_t$ the search history (incumbent
solutions and objective trajectory), $d_t$ the execution diagnostics
(e.g., calibration data, noise estimates, and feasible-sample rates), and
$r_t$ the remaining resource budget. The action
\begin{equation}
  a_t = \bigl(f_t,\; g_t,\; e_t,\; q_t,\; \theta_t,\; b_t\bigr)
  \label{eq:action}
\end{equation}
jointly specifies the formulation $f_t$ (including penalty weights), the
decomposition or subproblem $g_t$, the encoding $e_t$, the solver choice
$q_t$ (quantum or classical, and which algorithm), the solver parameters
$\theta_t$ (e.g., QAOA angles), and the backend and sampling budget $b_t$. A
policy $\pi(a_t \mid s_t)$ is then evaluated by the system-level objective
\begin{equation}
  J(\pi) = \E_{x \sim \Dist,\; \tau \sim \pi}
  \bigl[-\lambda_1 \Delta - \lambda_2 T - \lambda_3 \NQPU
        - \lambda_4 \Nshots - \lambda_5 V\bigr],
  \qquad
  \pi^{\star} \in \arg\max_{\pi} J(\pi),
  \label{eq:objective}
\end{equation}
where the expectation is over instances drawn from a distribution $\Dist$ and
over solver trajectories $\tau$ generated by $\pi$; $\Delta$ is the
optimality gap of the returned solution, $T$ the total wall-clock time,
$\NQPU$ the number of QPU calls (or annealer submissions), $\Nshots$ the
total number of shots, and $V$ the residual constraint violation. The
weights $\lambda_1,\dots,\lambda_5 \ge 0$ encode the relative price of
quality, time, quantum access, sampling, and infeasibility.

In this notation, a static pipeline is a fixed action sequence, and most
AI-for-QO methods learn a single component of $a_t$ (for example, $\theta_t$)
while holding the others fixed. Co-optimization either learns several
interacting components jointly or adapts them sequentially as the solver state
evolves, including repeated state-dependent allocation between classical and
quantum actions.

\subsection{Solver Selection as a Precursor to Co-Optimization}
\label{sec:selection}

Solver-selection methods \citep{moussa2020,volpe2024}, discussed in
\cref{sec:hardware}, do not satisfy our definition of co-optimization,
because only the solver-choice decision $q_t$ is learned; we therefore code
them as AI-for-QO. We nevertheless treat them as important precursors. They
make quantum computation one option in an algorithm portfolio, selected
conditionally on instance features, and so introduce ex-ante allocation
between classical and quantum resources. The systems below go further by
learning several coupled decisions, or by making the allocation decision
repeatedly during the search.

\subsection{Learning Problem Structure and Optimization Dynamics}
\label{sec:structure}

MLQAOA, graph-shrinking methods, and learned decomposition extend
orchestration upstream by learning the problem structure presented to the QPU
\citep{bach2024,zheng2026}. Hybrid CVRP methods by Sharma and Lau use
reinforcement learning to adapt augmented-Lagrangian or Lagrangian controls
while bounded quantum subproblems are solved inside an OR decomposition
\citep{sharma2025hybrid}. A later architecture adds contextual selection of
quantum execution configurations \citep{sharma2026cvrp}. The decisions are
coupled: multiplier updates alter the subproblem structure ($g_t$), which
changes quantum difficulty, which in turn should affect execution decisions
($b_t$).

\subsection{Selective Quantum Local Search}
\label{sec:alns}

The study of Moosavi and Farooq~\citep{moosavi2026} is the clearest example of online
selective invocation in the screened corpus: during the search, the
controller can reject quantum repair and select a classical operator instead. A deep
Q-network chooses among classical repair operators and shallow quantum
circuits using features of the current repair state and predicted hardware
reliability. Quantum actions are masked when the reduced problem is too large
or unreliable.

In their experiments, at least one quantum action is admissible in only about
16\% of reduced repair contexts, and classical repair is better on average
even within that subset. Quantum-enabled repair nevertheless improves final
outcomes in some matched-budget regimes. These results support selective
quantum invocation, not general quantum superiority: quantum repair acts as a
scarce heuristic resource whose use is determined by the learned controller.

\subsection{Resource-Aware Automation and Policy Search}
\label{sec:agentic}

The QuaST Decision Tree \citep{poggel2026} provides a modular automation
architecture spanning modeling, encoding, algorithm selection,
hyperparameters, execution, and interpretation. Its recommendation layer
estimates whether candidate variational configurations are likely to exceed a
classical disadvantage boundary and can route infeasible configurations away
from quantum execution. Related resource-control work includes
reinforcement-learning-based adaptive shot allocation for recursive QAOA
\citep{leekim2026} and joint learned graph partitioning with QAOA
initialization \citep{zheng2026}. These systems embody an important design
principle: an intelligent quantum optimizer must also know when \emph{not} to
use quantum computation.

AutoQResearch \citep{sharma2026autoq} moves the abstraction one level higher.
An LLM-based agent proposes modifications within a constrained solver-policy
space, while a fixed evaluation harness determines whether the changes
improve performance. A staged scout--promote--confirm procedure separates
rapid exploration from stronger confirmation. The object of optimization is
therefore a conditional solver policy $\pi$ rather than only a parameter
vector. Current evidence should not be described as autonomous algorithm
discovery, because the search space and evaluation protocol remain externally
specified, but it is a genuine transition toward closed-loop solver-policy
search.

\subsection{Boundary Cases}
\label{sec:boundary}

We reserve the co-optimization label for methods that learn either several
coupled decision dimensions or repeated, state-dependent allocation between
heterogeneous solver actions; selective quantum local search
(\cref{sec:alns}) qualifies under the second criterion. DQAOA-GPT
\citep{kim2026dqaoa}, for example, generates circuits for decomposed
subproblems, but the decomposition itself is random rather than learned
jointly with circuit generation. We therefore classify it as AI-assisted
circuit synthesis inside a distributed framework rather than as
co-optimization.

\Cref{sec:aiqo,sec:qai} and this section are organized by what is learned.
The next section reorganizes the same literature by problem family, which
shows where the methods have been tested and where the gaps lie.

\section{Problem-Centered Synthesis}
\label{sec:problems}

\Cref{tab:problems} summarizes, for each major problem family, why it matters,
the dominant AI--quantum approaches, the main bottleneck, and the principal
research opportunity. The subsections below discuss each family in turn.

\begin{table}[t]
  \caption{Problem-centered synthesis of AI--quantum approaches by problem
    family.}
  \label{tab:problems}
  \small
  \begin{tabularx}{\linewidth}{@{}P{1.9cm}LLLL@{}}
    \toprule
    \textbf{Problem family} & \textbf{Why it matters}
      & \textbf{Dominant AI--quantum approaches} & \textbf{Main bottleneck}
      & \textbf{Research opportunity} \\
    \midrule
    MaxCut & clean benchmark; extensive QAOA literature
      & parameter transfer, BO, QAS, error mitigation
      & benchmark saturation; no hard constraints
      & cross-topology and hardware transfer \\
    MIS / clique & constraint-sensitive graph structure
      & recursive policies, QAOA/QRAO, learned graph models
      & feasibility--compression trade-offs
      & adaptive encoding and resource allocation \\
    TSP & canonical sequential CO
      & equivariant QRL, quantum attention
      & all-to-all interactions; shot sensitivity
      & hardware-aware learned routing policies \\
    CVRP / VRP & realistic constrained routing
      & decomposition, RL control, quantum local repair
      & $O(n^2)$ formulations; capacity and time windows
      & selective quantum subproblems and repair \\
    Knapsack / MDKP & compact but constraint-rich
      & QRAO, slack-free penalties, Hamiltonian QRL
      & inequality encoding; feasible sampling
      & adaptive penalty and compression selection \\
    QAP / assignment & bridge between QUBO and learned prediction
      & QNNs, end-to-end quantum learning
      & dense coupling; sparse feasible states
      & amortized prediction vs.\ per-instance solve \\
    Scheduling & high industrial relevance
      & QAS, QUBO/VQA hybrids
      & temporal and resource constraints
      & stronger orchestration benchmarks \\
    \bottomrule
  \end{tabularx}
  \par\smallskip
  {\footnotesize\raggedright Abbreviations: BO, Bayesian optimization; CO,
    combinatorial optimization; QNN, quantum neural network; $n$, number of
    customers or cities.\par}
\end{table}

\subsection{MaxCut and MIS}
\label{sec:maxcut}

MaxCut remains the dominant methodological testbed because it has a simple
Ising form and abundant benchmarks. It has enabled controlled studies of
parameter prediction, transfer, Bayesian optimization, architecture search,
and noise mitigation. Its weakness is precisely that it under-represents the
feasibility and modeling difficulties that dominate constrained OR problems.
MIS and clique problems are more revealing because feasibility, recursion,
and compression become central \citep{bravyi2020,patel2023,duneau2026}.

\subsection{TSP and VRP}
\label{sec:routing}

Routing makes the case against monolithic quantum optimization. Standard
formulations require $O(n^2)$ binary variables for $n$ cities or customers
before capacity, fleet, or time-window constraints are added. The strongest
AI--quantum routing work therefore emphasizes sequential policies,
decomposition, learned assignment, local repair, and resource-aware execution
\citep{skolik2023,sharma2026cvrp,moosavi2026}. The emerging architecture is
modular rather than monolithic.

\subsection{Knapsack and MDKP}
\label{sec:knapsack}

Knapsack-type models expose inequality-constraint issues while remaining
compact enough for controlled experiments. They support comparisons among
slack variables, custom penalties, QRAO, qubit-efficient encodings, and
learned constraint control
\citep{sharma2024qrao,leelau2025slack,leelau2026cvar,kruse2024}. They also
reveal a key systems trade-off: fewer qubits do not automatically imply lower
total computational cost if measurement, decoding, or repair becomes harder.

\subsection{QAP, Assignment, and Contextual Optimization}
\label{sec:qap}

Assignment connects direct QUBO optimization with quantum machine learning.
The same problem family can be attacked per instance with a variational
optimizer or amortized across instances with a trained quantum predictor
\citep{ye2023,leekwon2026}. This makes it especially useful for separating
optimization advantage from learning advantage.

\subsection{Scheduling and Resource Allocation}
\label{sec:scheduling}

Scheduling is still under-represented relative to graph optimization, but it
is important for external validity. Differentiable quantum architecture
search has already been studied for job-shop scheduling \citep{sun2024}, and
the broader quantum-scheduling literature is now large enough to support
dedicated reviews \citep{werner2026}. Temporal precedence, resource coupling,
and feasibility structure differ sharply from MaxCut, making scheduling a
natural stress test for claims that an AI--quantum method generalizes across
combinatorial problem classes.

Across the problem families, the same pattern recurs: methods are developed
and validated on small instances, often of unconstrained problems, and the
constrained problems with the greatest operational value are where the
evidence is thinnest. This raises the question of how strong the evidence
actually is, which the next section addresses.

\section{Evidence Maturity: What Has Actually Been Demonstrated?}
\label{sec:maturity}

\Cref{sec:aiqo,sec:qai,sec:coopt,sec:problems} described what each line of
work does. This section asks what it has demonstrated. It grades each of the
\nCore{} non-survey studies in the evidence core on the M0--M5 hierarchy, compares the three paradigms,
and then explains why the evidence stalls where it does.

The literature uses terms such as \emph{scaling}, \emph{generalization},
\emph{hardware demonstration}, and \emph{advantage} inconsistently. We use the
M0--M5 hierarchy in \cref{tab:maturity} only to record the strongest evidence
demonstrated by a study. It is not a ranking of scientific quality: a
theoretical paper may remain at M1, while an engineering integration study may
reach M3 or M4. Where the evidence straddles two levels (for example, a
hardware result for only part of the workflow), the matrix records a range
such as M3--M4.

\begin{table}[t]
  \caption{The M0--M5 evidence-maturity hierarchy. Each level records what a
    study demonstrates, not its scientific quality.}
  \label{tab:maturity}
  \small
  \begin{tabularx}{\linewidth}{@{}P{0.9cm}LLL@{}}
    \toprule
    \textbf{Level} & \textbf{Evidence requirement} & \textbf{Establishes}
      & \textbf{Does not establish} \\
    \midrule
    M0 & conceptual mechanism or architecture & plausibility
       & empirical effectiveness \\
    M1 & ideal statevector or exact simulation & proof of concept
       & noise robustness or scaling \\
    M2 & held-out, cross-size, cross-topology, or cross-problem tests
       & some generalization & hardware viability \\
    M3 & realistic noise or physical-QPU validation of a material component
       & hardware relevance of that component & end-to-end system benefit \\
    M4 & complete hybrid workflow with physical quantum execution
       & operational hardware integration & practical superiority \\
    M5 & matched-resource comparison against strong classical and
         static-hybrid alternatives
       & practical quantum advantage in the tested regime
       & asymptotic quantum advantage, unless separately proven \\
    \bottomrule
  \end{tabularx}
\end{table}

\begin{figure}[t]
  \centering
  \input{figures/fig3_autonomy_evidence}
  \caption{Scope of learned solver control versus the strongest quantum
    evidence reported by representative studies. Vertical positions are the maturity
    levels coded in the evidence matrix (\cref{app:fields}); translucent bars mark studies
    coded as a range (e.g., M3--M4), with the marker at the midpoint.
    Horizontal positions are qualitative. Marker shape and colour indicate
    the paradigm.}
  \label{fig:autonomy}
\end{figure}
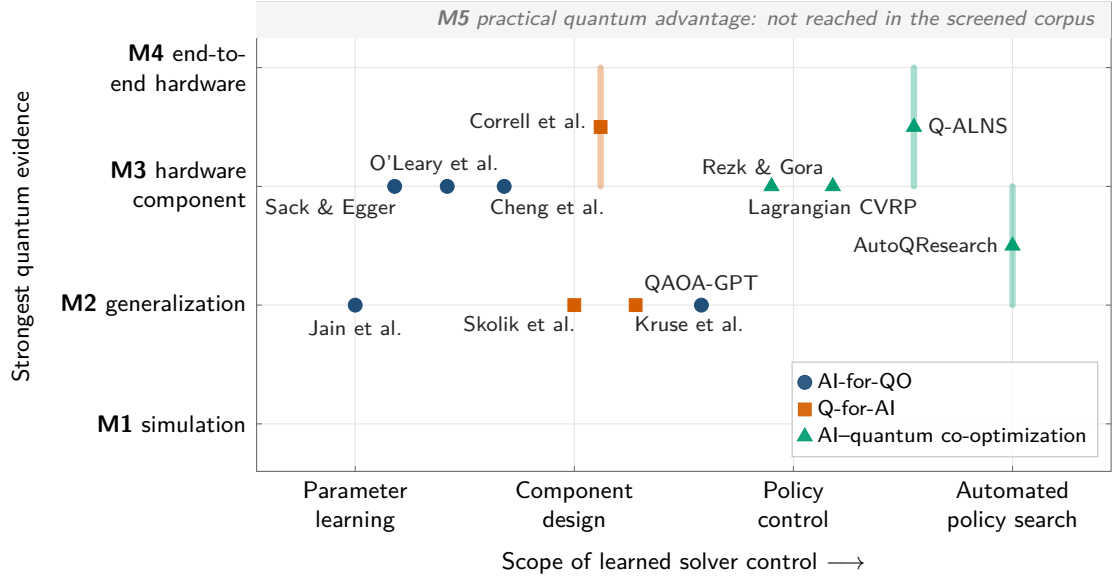

\subsection{The Asymmetry of the Literature}
\label{sec:asymmetry}

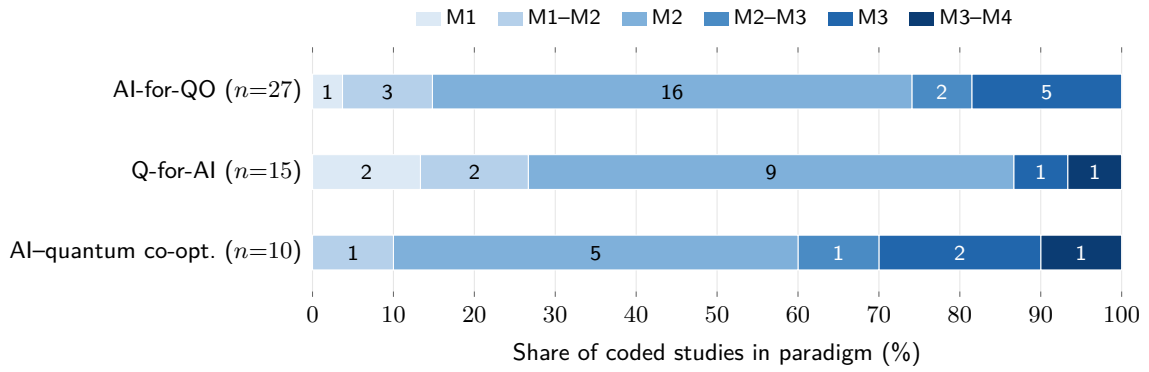
\begin{figure}[t]
  \centering
  \input{figures/fig4_maturity_by_paradigm}
  \caption{Distribution of evidence maturity within each paradigm for the
    \nParadigm{} paradigm studies in the evidence core, using the coding described in \cref{app:fields}. Bar segments show the
    share of studies at each level; numbers inside segments are study counts.
    Background reviews (10), structural precursors (4), and one benchmark
    study are omitted. No study receives an unqualified M4 or M5
    assignment; some studies straddle M3--M4.}
  \label{fig:maturity}
\end{figure}

\Cref{fig:autonomy} places representative studies by the scope of their AI
component and the strength of their quantum evidence, and \cref{fig:maturity}
summarizes the maturity distribution of the \nParadigm{} studies in the three
paradigms (the \nCore{} non-survey core studies less the four structural
precursors and one benchmark).
Three observations stand out.

The largest body of hardware-grounded evidence is on the AI-for-QO side.
Studies have reduced QPU evaluations, transferred parameters, mitigated noise,
adapted annealing schedules \citep{finzgar2024,leclerc2025}, and incorporated
hardware-aware execution. Seven of the \nAIforQO{} coded AI-for-QO studies
reach at least M2--M3. These results are practically relevant even without quantum
advantage because they reduce QPU evaluations, improve robustness, or lower
execution overhead. Their natural comparator is the unassisted quantum
routine (random or heuristic initialization, a default schedule), and
\nAIforQONoBase{} of the \nAIforQO{} AI-for-QO studies include no classical
solver at all. That is appropriate for the question they ask, but it means
they show that learning makes quantum optimization cheaper, not that the
resulting pipeline is competitive with classical optimization.

Q-for-AI, in contrast, is less mature. Existing work shows that quantum
policies can encode symmetry, reduce parameter counts, and generalize across
selected instance distributions, and 13 of the 15 coded studies remain at
M1--M2. Classical learned baselines are common: \nQforAILearned{} of the
\nQforAI{} Q-for-AI studies include one. The gap
lies elsewhere. Only \nQforAIQAblYes{} of the \nQforAIQAblApp{} studies where
an ablation applies replace the quantum layer with a classical layer of
matched size, so the quantum contribution is often confounded with
differences in architecture or parameter count; training and inference
budgets are rarely matched; and only \nQforAIHardware{} studies execute on
physical hardware, among them the partially hardware-executed logistics model
of Correll et al.~\citep{correll2022}. The evidence is therefore stronger for
architecture and inductive bias than for practical superiority.

Co-optimization is the smallest and most recent category. Credible examples
now exist in learned decomposition, shot allocation, selective local quantum
search, and policy-level automation. Five of the \nCoopt{} coded studies
contain some physical-hardware evidence, although only four receive an M2--M3
or higher maturity assignment, because in the fifth the hardware runs are not
material to the claimed contribution. However, the
screened corpus does not contain a broadly convincing M5 result in any
paradigm.

\subsection{Why the Evidence Stalls: Scaling as a Systems Problem}
\label{sec:scaling}

The asymmetry above has a common cause. The experiments behind it are small:
among the \nQubitsStated{} paradigm studies that state a qubit count, the
median of the largest reported experiment is \medQubits{} qubits,
\nQubitsLeTwenty{} stay at 20 qubits or fewer, and only \nQubitsGtHundred{}
exceed 100; \nHardware{} of the \nParadigm{} paradigm studies execute any part
of the workflow on physical hardware. It is tempting
to read this as a qubit-count problem that larger devices will solve. The
reviewed studies suggest otherwise. They expose several bottlenecks besides
qubit count: two-qubit gate counts and transpiled depth, shot budgets,
feasible-state density, penalty conditioning, QPU calls and latency, classical
repair, and the trainability of the variational objective
(\cref{sec:shots,sec:trainability}).

These quantities interact, especially for constrained problems. A
qubit-efficient encoding can make a larger instance representable while
making decoding or repair harder; a slack-free penalty can save auxiliary
qubits while changing the objective landscape; and decomposition can cap
quantum width while increasing coordination overhead
(\cref{sec:formulation,sec:decomposition}). Early hardware QAOA experiments
already showed performance degrading as problem graphs departed from the
native device topology \citep{harrigan2021}. Real-hardware benchmarking across
MDKP, MIS, QAP, and market-share problems illustrates the point
\citep{sharma2026hw}: qubit-efficient methods extend the range of executable
instances, yet dense constraints and compiled two-qubit gate counts push
circuits into a noise-dominated regime, and in QAP feasible states are so
sparse that successful circuit execution does not imply useful feasible
sampling.

Two consequences follow for the rest of the review. First, moving a result
from M3 to M4 or M5 requires accounting for the whole system rather than the
quantum routine alone, so benchmarks must report system-level resources and
isolate the quantum contribution (\cref{sec:benchmarking}). Second, when the
cost of a quantum call depends on so many interacting factors, deciding
whether a call is worth making becomes a central design problem: the problem
of selective quantum use taken up in \cref{sec:when-not}.

\section{Benchmarking Principles for \AIxQ{} Optimization}
\label{sec:benchmarking}

The principles below follow from the gaps identified in
\cref{sec:asymmetry,sec:scaling}. Each states what an experiment should
report and, where the evidence fields allow it, how far current practice
falls short.

\subsection{Report System-Level Resources}
\label{sec:resources}

At a minimum, experiments should report solution quality, feasibility, total
runtime, QPU calls, shots, physical and logical qubits, transpiled depth,
two-qubit gate count, training cost, and the hardware and noise regime. Cloud
latency and monetary cost should be included when material. Energy should be
reported only when directly measured or credibly estimated.
\Cref{tab:reporting} consolidates these expectations into a checklist.
Current practice falls well short of it. Among the \nResCoded{} paradigm
studies, \nShotsReported{} state the shot or sample
budget, \nExactExp{} use exact expectation values, and \nShotsNotReported{}
do not report shots at all; only \nCallsReported{} report the number of
quantum evaluations or QPU calls, and \nBothResources{} report both.

\subsection{Separate Hybrid-System Benefit from Quantum Contribution}
\label{sec:separate}

A hybrid solver can improve on a baseline even when its quantum component
contributes little. Evaluation should therefore answer two separate
questions: does the complete system improve on relevant baselines, and does
the quantum component make a measurable marginal contribution? The first does
not imply the second.

\subsection{Use Strong Baselines and Matched Decision Budgets}
\label{sec:baselines}

A QRL model should be compared against modern classical learned optimization,
not only against another quantum model. A routing system should include
credible OR and heuristic baselines, such as state-of-the-art exact solvers
(e.g., Gurobi) and highly tuned heuristics (e.g., LKH-3). If a quantum repair
receives additional sampling budget, the classical alternative should receive
a meaningful matched compute or evaluation budget. Ablations should isolate
the contributions of the learned component and of the quantum component
separately. Among the \nParadigm{} paradigm studies, \nNoBase{} include
no classical baseline of any kind, \nStrongBase{} include a strong heuristic
or exact solver as a competitor, and \nLearnedBase{} include a classical
learned baseline. Where a quantum ablation applies, \nQAblYes{} of
\nQAblApplicable{} studies run a full ablation, \nQAblPartial{} a partial one,
and \nQAblNo{} none.

\subsection{Account for Amortization and Generalization}
\label{sec:amortize}

Learning cost matters when a model is reused across instances. Papers should
report the break-even number of instances $N^{\star}$ in
\cref{eq:breakeven} where it can be estimated. Generalization claims should
also state the tested transition explicitly: new instances, larger sizes, new
topologies, new problem families, new noise regimes, or new hardware.

\begin{table}[t]
  \caption{Minimum reporting checklist for \AIxQ{} optimization experiments.}
  \label{tab:reporting}
  \small
  \begin{tabularx}{\linewidth}{@{}P{3.2cm}L@{}}
    \toprule
    \textbf{Dimension} & \textbf{Minimum reporting expectation} \\
    \midrule
    Solution quality & gap or approximation ratio; best-case and
      distributional performance \\
    Feasibility & feasible-sample rate; constraint violation; repair \\
    Quantum resources & qubits; two-qubit gates; depth \\
    Sampling & shots per call; total shots; allocation policy \\
    Invocation cost & QPU calls or annealer submissions \\
    Time & training; classical compute; QPU execution; queue time where
      material \\
    Learning & train/validation/test split; number of training instances;
      model size; break-even $N^{\star}$ \\
    Generalization & instance; size; topology; problem; noise; hardware \\
    Baselines & strong OR, heuristic, and ML baselines; static-hybrid
      ablations \\
    Hardware & device; compilation; calibration and noise; mitigation \\
    Reproducibility & code; seeds; instances or data; hyperparameters \\
    Cost & monetary and/or energy cost, only where credibly measured \\
    \bottomrule
  \end{tabularx}
\end{table}

\section{Open Problems and Outlook}
\label{sec:agenda}

The preceding sections identify where the evidence is thin. This section
turns each gap into an open problem. For each problem we state the gap and
where it is documented, a concrete question, and the kind of study, expressed
in terms of the evidence hierarchy of \cref{tab:maturity}, that would settle
it. We close with an outlook on how the field has developed and where it is
heading.

\subsection{Learning Formulations, Not Only Parameters}
\label{sec:learn-form}

\paragraph{Gap.} Most AI-for-QO work targets QAOA angles
(\cref{sec:params,sec:query}), yet on constrained problems the formulation,
including penalty weights, slack encodings, compression, and decomposition,
often determines performance before any angle is chosen
(\cref{sec:formulation,sec:scaling}).

\paragraph{Question.} Can a learned controller that selects the formulation
jointly with the circuit parameters improve the feasible-sample rate and the
optimality gap at fixed qubit and shot budgets, compared with tuning the
parameters of a fixed formulation?

\paragraph{Evidence that would settle it.} Cross-problem tests at M2, followed
by hardware validation at M3, with an ablation that freezes the formulation so
that the contribution of formulation learning can be isolated.

\subsection{Learning When Not to Use a QPU}
\label{sec:when-not}

\paragraph{Gap.} This is the thread that runs through the review. The cost
of a quantum call depends on many interacting factors (\cref{sec:scaling}).
Instance-level solver selection already allows ex-ante rejection of quantum
execution \citep{moussa2020,volpe2024}. Much rarer is state-dependent
selective invocation during an ongoing hybrid search; the clearest example in
the screened corpus is a single study (\cref{sec:alns}). Most pipelines still
send every eligible subproblem to the QPU.

\paragraph{Question.} Can a controller predict, before invoking the QPU,
whether the expected marginal utility of quantum execution is positive?
Near-term systems should treat quantum calls as costly actions. At a decision
state $s$, the relevant quantity is the expected marginal value of quantum
execution relative to a classical alternative, after accounting for shots,
latency, and reliability:
\begin{equation}
  \Delta U_{\aQ}(s) =
  \E\bigl[\,U(s') \mid s,\, a = \aQ\,\bigr]
  - \E\bigl[\,U(s') \mid s,\, a = \aC\,\bigr]
  - c_{\aQ}(s),
  \label{eq:marginal}
\end{equation}
where $a=\aQ$ and $a=\aC$ denote the quantum and classical actions, $s'$ is
the resulting solver state, $U$ is a utility consistent with the objective in
\cref{eq:objective}, and $c_{\aQ}(s)\ge 0$ is the incremental cost of quantum
execution (shots, latency, queueing, and reliability risk). The quantum
action is preferred only when $\Delta U_{\aQ}(s) > 0$, and the question is
whether this sign can be predicted reliably from cheap diagnostics, such as
subproblem size, feasible-state density, and calibration data.

\paragraph{Evidence that would settle it.} An M4 workflow in which the learned
invocation policy is compared, under matched total budgets, against both an
always-classical and an always-quantum variant. The study should report how
often the quantum action is chosen and its marginal contribution to the final
result (\cref{sec:separate}).

\subsection{Closing the Gap Between Model and System Generalization}
\label{sec:model-system}

\paragraph{Gap.} Thirteen of the 15 coded Q-for-AI studies remain at M1--M2
(\cref{fig:maturity}), and transfer that holds in ideal simulation can degrade
under finite shots, transpilation, and noise (\cref{sec:generalization}).

\paragraph{Question.} Do the parameter savings of symmetry-aware quantum
policies survive on hardware, and do they outperform classical neural
combinatorial optimizers with a matched parameter count and training compute?

\paragraph{Evidence that would settle it.} Cross-size tests on physical
hardware (M3) against a matched classical neural baseline, reporting the
training cost of both models and the break-even point of
\cref{eq:breakeven}.

\subsection{Moving Beyond MaxCut}
\label{sec:beyond-maxcut}

\paragraph{Gap.} Much of the controlled evidence comes from MaxCut, which has
no hard constraints (\cref{sec:maxcut}). Routing, assignment, scheduling, and
knapsack problems bring sparse feasibility, repair, and stochasticity, and
are where operational value would lie (\cref{tab:problems}).

\paragraph{Question.} Do parameter-transfer and initialization methods
developed on MaxCut carry over to problems with sparse feasible subspaces?
More generally, which constraint structures admit useful quantum subproblems,
when is compression preferable to decomposition, and which diagnostics
predict a useful quantum repair?

\paragraph{Evidence that would settle it.} Cross-problem studies at M2 or
above that report feasibility metrics alongside solution quality, so that
claims of generality are tested against problem classes with different
constraint structure.

\subsection{Shared Environments for Evaluating Orchestration}
\label{sec:orch-bench}

\paragraph{Gap.} No study in the screened corpus reaches M5. One reason is
that there is no shared environment in which hybrid policies can be compared
against strong classical solvers under matched resources; each study builds
its own (\cref{sec:asymmetry,sec:benchmarking}).

\paragraph{Question.} Can the community agree on benchmark environments that
expose several problem families, strong classical solvers, multiple quantum
algorithms and backends (including noisy emulators), and explicit resource
budgets, so that the learning task is the orchestration policy $\pi$ of
\cref{eq:objective} rather than a single circuit?

\paragraph{Evidence that would settle it.} Such environments are a
precondition for M5 claims. Results reported in them should follow the
checklist in \cref{tab:reporting}.

\subsection{Trustworthy Automated Experimentation}
\label{sec:trustworthy}

\paragraph{Gap.} LLM-guided search over solver policies has begun to appear
(\cref{sec:agentic}), but automated search can overfit its evaluation harness
in ways that are hard to detect.

\paragraph{Question.} What controls are needed before results produced by
automated policy search can be trusted? At a minimum: fixed evaluation harnesses,
separate search and held-out instance sets, explicit search budgets, replay
of winning candidates, independent confirmation, provenance for every solver
modification, and ablations that identify which change produced the gain.

\paragraph{Evidence that would settle it.} At minimum M2 evidence on held-out
instances, followed by independent replication, with the search budget
reported alongside the result; where the searched policy controls physical
hardware, M3 validation should also be required.

\subsection{Outlook}
\label{sec:outlook}

Read conceptually rather than chronologically, the literature spans four
progressively broader questions. The first is how feasibility can be
preserved and how constrained problems can be represented without excessive
penalties or auxiliary variables, through constraint-preserving mixers \citep{hadfield2019},
feasibility-preserving encodings \citep{xie2024}, qubit-efficient relaxations
\citep{fuller2021,sharma2024qrao}, and slack-free penalties
\citep{leelau2025slack}. The second comes from hardware studies, which show
that fitting a model into the available qubits is not sufficient, because noise, compilation, and
sparse feasible states determine whether execution is useful
\citep{sack2024,sharma2026hw}; this motivated decomposition and learned
preprocessing \citep{bach2024,rezk2026}. The third concerns learned
optimization controls and resource allocation: solver selection
\citep{moussa2020,volpe2024}, adaptive penalty, multiplier, and repair
choices \citep{ayanzadeh2020,sharma2025hybrid,moosavi2026}, and automation
that rejects unfavorable quantum configurations \citep{poggel2026}. The broadest asks
how to learn policies conditioned on the current solver state,
through adaptive shot allocation \citep{leekim2026}, joint partitioning and
initialization \citep{zheng2026}, and LLM-guided circuit generation and
policy search \citep{kim2026dqaoa,sharma2026autoq}. The underlying question
is therefore no longer only ``which quantum algorithm?'' but ``which
formulation, subproblem, solver, and resource should be used at this point
in the search?'' The open problems above are the steps needed to answer it
with evidence at the level of M4 and M5 rather than M1 and M2.

\section{Conclusion}
\label{sec:conclusion}

Quantum combinatorial optimization is no longer only a question of choosing a
quantum algorithm for a fixed problem representation. AI adds decisions about
how the algorithm is initialized and configured; quantum machine learning asks
whether quantum models can improve learned optimizers; and co-optimization
extends the problem to the allocation of classical and quantum resources over
time.

The three directions are not equally mature. AI-for-QO already has credible
evidence in parameter transfer, query-efficient search, circuit synthesis,
graph reduction, penalty adaptation, noise mitigation, and hardware-aware
execution. Q-for-AI remains more exploratory: results support architectural
feasibility, symmetry, compact parameterization, and limited generalization,
but rarely practical superiority over strong classical learned optimization.
Co-optimization is beginning to connect solver selection, decomposition, shot
allocation, local quantum search, hardware-aware control, and policy search.

The evidence in all three is shallower than the volume of work suggests,
and its limits are set by the whole system rather than by qubit counts
alone. For near-term systems, therefore, selective QPU use is a more
realistic objective than routing every eligible subproblem to quantum
hardware. A well-designed hybrid
optimizer should be able to recognize when a QPU is unlikely to help and
choose a classical alternative. Progress beyond QAOA will therefore depend on
whether hybrid systems can identify the subproblems and execution regimes in
which the marginal value of quantum computation exceeds its additional cost.
Seen this way, the field is moving along a trajectory from
\begin{center}
  \small
  quantum algorithm
  $\;\longrightarrow\;$ AI-assisted quantum algorithm
  $\;\longrightarrow\;$ adaptive hybrid solver
  $\;\longrightarrow\;$ heterogeneous solver policy.
\end{center}

\section*{Data Availability}
\Cref{app:fields} defines the fields used to code each study and the rules
for assigning maturity levels. The coded data, with one row per paper for all
119 papers in the reference corpus and detailed coding for the 67 papers in
the evidence core, is available as a spreadsheet accompanying this article
(\texttt{Beyond\_QAOA\_\allowbreak Supplementary\_\allowbreak Evidence\_\allowbreak Matrix.xlsx}).
Coding follows the two tiers of \cref{app:fields}. Tier~1 descriptive
fields are recorded for core papers where the paper provides them and support
the qualitative synthesis and maturity assignments. The \emph{Evidence Coding}
sheet holds the Tier~2 evidence fields (F1--F4, fixed categories) and the
maturity level for each of the \nCore{} non-survey core studies, with the
location in the paper that supports each value; background surveys are not
Tier~2 coded. The \emph{Counts} sheet computes every count reported in the
text and figures from those Tier~2 entries, so all reported counts and maturity
distributions can be reproduced from it.

\section*{Acknowledgments}
The author used an AI assistant (Claude, Anthropic) for language editing,
LaTeX preparation, and first-pass extraction of evidence-matrix entries. All
content was reviewed and verified by the author.

\bibliographystyle{quantum}
\bibliography{references}

\appendix
\section{Evidence-Matrix Coding Scheme}
\label{app:fields}

\Cref{tab:fields} specifies the coding scheme for the evidence matrix. Each
core study occupies one row. The scheme has two tiers. Tier~1 fields are
descriptive and are recorded where the paper provides them; they support the
qualitative synthesis and the M0--M5 maturity assignments. Tier~2 fields
(F1--F4) use fixed categories and are coded for every non-survey core study,
so that every count in the text can be recomputed. For Tier~2, \emph{not
reported} means that the paper was read and the information is absent, and
\emph{not checked} means that the full text was unavailable; studies that are
not checked on a field are excluded from that field's denominator. The coded
matrix accompanies the article as a spreadsheet (see Data Availability).

{\small
\begin{xltabular}{\linewidth}{@{}P{2.8cm}LP{3.8cm}@{}}
  \caption{Evidence-matrix fields and coding rules.}\label{tab:fields}\\
  \toprule
  \textbf{Field} & \textbf{Definition / coding rule}
    & \textbf{Typical values / notes} \\
  \midrule
  \endfirsthead
  \multicolumn{3}{@{}l}{\footnotesize\itshape \Cref{tab:fields} continued}\\
  \toprule
  \textbf{Field} & \textbf{Definition / coding rule}
    & \textbf{Typical values / notes} \\
  \midrule
  \endhead
  \midrule
  \multicolumn{3}{r@{}}{\footnotesize\itshape continued on next page}\\
  \endfoot
  \bottomrule
  \endlastfoot
  \multicolumn{3}{@{}l}{\textit{Tier 1: descriptive fields (recorded where available)}}\\[2pt]
  Citation key & Short unique identifier used consistently in the review and
    the supplementary matrix. & e.g., Jai21, Ole25, Sha26c \\
  Year & Publication year; for preprints, the version year used at the
    review cut-off. & 2019--2026 in current corpus \\
  Problem family & Primary combinatorial optimization problem(s), e.g.,
    MaxCut, MIS, TSP, CVRP/VRP, MDKP/knapsack, QAP/assignment, scheduling.
    & One or more problem tags \\
  Paradigm & AI-for-QO, Q-for-AI, or AI--Quantum Co-Optimization. Boundary
    cases may receive a secondary tag, but one primary classification is
    assigned. & AI-for-QO / Q-for-AI / AI--Quantum Co-Optimization \\
  AI method & Learning or AI mechanism used, e.g., GNN, reinforcement
    learning, Bayesian optimization, surrogate learning, transformer/LLM,
    diffusion model, classifier, contextual bandit.
    & GNN, RL, BO, surrogate, LLM, bandit, etc. \\
  Quantum method & Quantum component, e.g., QAOA, VQE, QRAO, quantum
    annealing, neutral-atom optimization, parameterized quantum circuit/QNN,
    equivariant quantum circuit. & QAOA, VQE, QRAO, annealer, PQC/QNN, etc.\\
  Pipeline intervention & Where AI intervenes: parameter initialization,
    query control, circuit/ansatz design, formulation/penalties,
    reduction/decomposition, sampling, error mitigation, hardware execution,
    solver selection, or solver-policy control.
    & One primary intervention; secondary tags allowed \\
  Learned object & What is learned or predicted: parameters, donor instance,
    circuit structure, penalty, decomposition, shot budget, backend, solver
    choice, repair action, or conditional solver policy.
    & Angles, circuit, penalty, partition, shots, backend, policy, etc. \\
  Training regime & Per-instance optimization, offline supervised learning,
    RL interaction, meta-learning, transfer learning, zero-/few-shot
    generation, or other training set-up.
    & Per-instance / amortized / transfer / zero-shot, etc. \\
  Test / generalization regime & Whether evaluation is in-distribution,
    held-out instance, cross-size, cross-topology, cross-problem,
    cross-noise, or cross-hardware.
    & Instance / size / topology / problem / noise / hardware \\
  Problem size & Number of vertices, variables, customers, jobs, or another
    natural problem-size measure used in the study.
    & Natural domain scale, with mapping to binary variables \\
  Execution environment & Ideal statevector, shot-based simulation, noisy
    simulation/emulation, or named physical device. & Device class is coded
    in F4c \\
  AI ablation & Whether the contribution of the AI component is isolated by
    removing or replacing it; summarize the relevant ablation.
    & Yes / No / Partial, with result \\
  Feasibility metric & Feasible-sample rate, violation magnitude, repair
    rate, or other measure of constraint satisfaction, if applicable.
    & N/A for unconstrained problems \\
  Solution-quality metric & Optimality gap, approximation ratio, objective
    value, regret, reward, or a problem-specific solution-quality measure.
    & Gap, approximation ratio, regret, objective, reward \\
  Runtime / cost & Training time, classical compute, QPU execution,
    wall-clock time, queue latency, cloud monetary cost, or energy where
    credibly measured.
    & Separate training, inference, and QPU time where possible \\
  Hardware robustness & Evidence under realistic noise, calibration
    variation, compilation, or physical hardware, including sensitivity
    across devices if tested.
    & Ideal only / noisy / physical QPU / multi-device \\
  Reproducibility & Availability of code, data or instance generator, seeds,
    hyperparameters, and sufficient implementation details.
    & Code / data / seeds / hyperparameters \\
  Claimed benefit & Benefit stated by the authors, recorded without
    endorsement. & Author-stated claim \\
  Evidence-supported conclusion & What the reported experiments actually
    establish after considering baselines, resource matching,
    generalization, and hardware conditions.
    & Reviewer-coded synthesis statement \\
  Maturity level & Highest M0--M5 evidence level directly supported:
    M0 conceptual; M1 ideal simulation; M2 generalization; M3 hardware or
    noisy component validation; M4 end-to-end hardware-grounded workflow;
    M5 practical quantum advantage under matched resources.
    & M0--M5; ranges (e.g., M2--M3) where evidence straddles levels \\
  \midrule
  \multicolumn{3}{@{}l}{\textit{Tier 2: evidence fields (fixed categories, coded for every non-survey core study)}}\\[2pt]
  F1 Strongest classical non-learned baseline & Strongest classical solver
    that does not learn, reported as a competitor on the same instances.
    Quantum-only comparators do not count. An exact solver used only to
    compute the optimum for an approximation ratio is not a competitor.
    & None / Simple (greedy, random, default local search) / Strong
    (Goemans--Williamson, tuned SA or tabu, LKH, OR-Tools) / Exact
    (MILP, branch and bound) \\
  F2 Classical learned baseline & A classical learned model (NN, GNN, RL,
    neural CO, classical surrogate) that solves the same task and is
    reported as a competitor. & Yes / No \\
  F3 Quantum ablation & Whether the quantum component is replaced by a
    classical counterpart and the results compared. Partial: compared, but
    without matched size or budget, or only in some experiments.
    & Yes / Partial / No / N/A (quantum routine is the object of study,
    e.g., parameter prediction, circuit design, schedules) \\
  F4a Max qubits & Largest qubit count in any experiment (logical qubits;
    problem variables for annealing). & Integer / not reported \\
  F4b Max qubits on hardware & Largest qubit count executed on a physical
    device. Device calibration data or noise models do not count.
    & Integer; 0 if simulation only \\
  F4c Hardware type & Device class used for physical execution.
    & Gate-based QPU / annealer / neutral-atom or analog / multiple / none \\
  F4d Shots & Whether the shot or sample budget per evaluation is stated.
    & Reported / exact expectation / not reported / not applicable \\
  F4e Quantum evaluations & Whether the number of objective evaluations,
    circuit executions, or QPU submissions is stated.
    & Reported / not reported / not applicable \\
\end{xltabular}
}

\end{document}

%% file: figures/fig0_paradigms.tex
\begin{tikzpicture}[
  font=\sffamily\small, >={Stealth[length=2.2mm]},
  every node/.append style={execute at begin node={\hyphenpenalty=10000\exhyphenpenalty=10000}},
  panel/.style={draw=#1, line width=0.9pt, rounded corners=4pt, minimum width=4.9cm, minimum height=8.3cm},
  ai/.style={draw=aiorange!90!black, fill=aiorange!18, rounded corners=3pt, align=center, minimum width=3.4cm, minimum height=0.95cm, line width=0.7pt},
  qu/.style={draw=qblue, fill=qblue!14, rounded corners=3pt, align=center, minimum width=3.4cm, minimum height=0.95cm, line width=0.7pt},
  cl/.style={draw=black!55, fill=black!5, rounded corners=3pt, align=center, minimum width=3.4cm, minimum height=0.95cm, line width=0.7pt},
  plain/.style={align=center, text=black!75},
  lab/.style={font=\sffamily\scriptsize, text=black!70, align=center},
  flow/.style={->, line width=0.8pt, black!70},
  fb/.style={->, line width=0.8pt, dashed, black!55},
]
\node[panel=aiqo] (PA) at (0,-0.2) {};
\node[font=\sffamily\bfseries, text=aiqo] at (0,3.55) {AI-for-QO};
\node[lab] at (0,3.05) {AI improves a quantum optimizer};
\node[plain] (iA) at (0,2.3) {problem instance};
\node[ai] (aA) at (0,1.05) {\textbf{learned model}\\ GNN, BO, RL};
\node[qu] (qA) at (0,-1.05) {\textbf{quantum optimizer}\\ QAOA, VQE, annealer};
\node[plain] (sA) at (0,-2.3) {solution};
\draw[flow] (iA) -- (aA);
\draw[flow, aiorange!80!black] (aA) -- node[lab, right, xshift=1mm, text=aiorange!70!black] {sets one\\decision} (qA);
\draw[flow] (qA) -- (sA);
\draw[fb] (qA.west) -- ++(-0.35,0) |- node[lab, pos=0.25, left, rotate=90, anchor=south] {evaluations} (aA.west);
\node[lab, text width=4.4cm, anchor=north] at (0,-2.95) {\textbf{Learned:} one stage, such as angles, penalties, shots or backend};

\begin{scope}[xshift=5.35cm]
\node[panel=qorange] (PB) at (0,-0.2) {};
\node[font=\sffamily\bfseries, text=qorange] at (0,3.55) {Q-for-AI};
\node[lab] at (0,3.05) {quantum layer inside the model};
\node[plain] (iB) at (0,2.3) {problem instance};
\node[draw=aiorange!90!black, fill=aiorange!8, rounded corners=3pt, line width=0.7pt, minimum width=3.9cm, minimum height=3.45cm] (pol) at (0,0) {};
\node[lab, text=aiorange!70!black, font=\sffamily\scriptsize\bfseries] at (0,1.45) {learned policy / predictor};
\node[cl, minimum height=0.6cm, minimum width=3.0cm] (enc) at (0,0.85) {classical encoder};
\node[qu, minimum height=0.75cm, minimum width=3.0cm] (pqc) at (0,-0.15) {\textbf{quantum layer} (PQC)};
\node[cl, minimum height=0.6cm, minimum width=3.0cm] (dec) at (0,-1.15) {classical decoder};
\draw[flow] (enc) -- (pqc); \draw[flow] (pqc) -- (dec);
\node[plain] (sB) at (0,-2.4) {action or solution};
\draw[flow] (iB) -- (pol.north);
\draw[flow] (pol.south) -- (sB);
\node[lab, text width=4.4cm, anchor=north] at (0,-2.95) {\textbf{Learned:} the model itself, with quantum layers replacing classical ones};
\end{scope}

\begin{scope}[xshift=10.7cm]
\node[panel=coopt] (PC) at (0,-0.2) {};
\node[font=\sffamily\bfseries, text=coopt] at (0,3.55) {AI--quantum co-opt.};
\node[lab] at (0,3.05) {AI allocates work across solvers};
\node[draw=coopt, fill=coopt!15, rounded corners=3pt, align=center, minimum width=3.3cm, minimum height=0.95cm, line width=0.7pt] (pi) at (0,2.05) {\textbf{solver policy} $\pi(a_t \mid s_t)$};
\node[qu, minimum width=1.75cm] (qC) at (-0.92,-0.05) {\textbf{quantum}\\ action};
\node[cl, minimum width=1.75cm] (cC) at (0.92,-0.05) {\textbf{classical}\\ action};
\draw[flow, coopt] (qC.north |- pi.south) -- (qC.north);
\draw[flow, coopt] (cC.north |- pi.south) -- (cC.north);
\node[lab, text width=1.75cm, font=\sffamily\tiny] at (0,1.05) {formulation, subproblem, solver, shots, backend};
\node[draw=black!55, fill=white, rounded corners=3pt, align=center, minimum width=3.3cm, minimum height=0.85cm, line width=0.7pt] (st) at (0,-1.6) {search state $s_t$\\[-1pt]{\scriptsize history, diagnostics, budget}};
\draw[flow] (qC.south) -- (qC.south |- st.north);
\draw[flow] (cC.south) -- (cC.south |- st.north);
\draw[fb] (st.west) -- ++(-0.3,0) |- node[lab, pos=0.25, left, rotate=90, anchor=south] {repeat} (pi.west);
\node[plain] (sC) at (0,-2.5) {solution};
\draw[flow] (st) -- (sC);
\node[lab, text width=4.4cm, anchor=north] at (0,-2.95) {\textbf{Learned:} several coupled decisions, or when quantum is worth calling};
\end{scope}
\end{tikzpicture}

%% file: figures/fig1_progression.tex
\begin{tikzpicture}[
  >=Stealth,
  stage/.style={draw=qblue!80!black, line width=0.6pt, rounded corners=3pt,
                minimum height=1.15cm, text width=2.05cm, align=center,
                font=\sffamily\footnotesize\bfseries, inner sep=3pt},
  sub/.style={font=\sffamily\scriptsize, text width=2.3cm, align=center,
              text=black!70},
]
  \def\step{2.72}
  \foreach \i/\name/\what [count=\k from 0] in {
      1/{Static quantum\\algorithm}/{QAOA, VQE,\\annealing},
      2/{AI-tuned quantum\\algorithm}/{parameters,\\schedules},
      3/{Learned hybrid\\optimizer}/{formulation,\\decomposition},
      4/{Adaptive solver\\policy}/{solver, shots,\\backend},
      5/{Heterogeneous\\solver policy\\(prospective)}/{CPU + GPU +\\QPU + annealer}} {
    \pgfmathtruncatemacro{\shade}{{8,20,36,66,88}[\k]}
    \ifnum\k>2
      \node[stage, fill=qblue!\shade, text=white] (s\i) at (\k*\step,0) {\name};
    \else
      \node[stage, fill=qblue!\shade] (s\i) at (\k*\step,0) {\name};
    \fi
    \node[sub, below=3pt of s\i] (w\i) {\what};
  }
  \foreach \a/\b in {1/2,2/3,3/4,4/5}
    \draw[->, line width=0.8pt, black!70] (s\a) -- (s\b);

  \coordinate (L) at ([yshift=-1.3cm]s1.south west);
  \coordinate (R) at ([yshift=-1.3cm]s5.south east);
  \shade[left color=qblue!10, right color=qblue!85!black]
        ([yshift=-2.2pt]L) rectangle ([yshift=2.2pt,xshift=-6pt]R);
  \fill[qblue!85!black] ([xshift=-6pt,yshift=5pt]R) -- (R) -- ([xshift=-6pt,yshift=-5pt]R) -- cycle;
  \node[font=\sffamily\footnotesize\itshape, below=3pt] at ($(L)!0.5!(R)$)
       {Increasing scope of the learned decision};
  \node[font=\sffamily\scriptsize, anchor=north west, text=black!60] at ([yshift=-3pt]L) {narrow};
  \node[font=\sffamily\scriptsize, anchor=north east, text=black!60] at ([yshift=-3pt]R) {broad};
\end{tikzpicture}

%% file: figures/fig_variational_loop.tex
\begin{tikzpicture}[
  >=Stealth,
  box/.style={draw, rounded corners=3pt, line width=0.6pt, align=center,
              font=\sffamily\footnotesize, inner sep=5pt, minimum height=1.25cm},
  cbox/.style={box, draw=black!65, fill=black!6, text width=3.1cm},
  qbox/.style={box, draw=qblue!80!black, fill=qblue!15, text width=3.3cm},
  lab/.style={font=\sffamily\scriptsize, text=black!80, align=center},
]
  \node[cbox] (opt) at (0,0)
    {\textbf{Classical optimizer}\\[1pt] proposes angles\\ $\boldsymbol{\theta}=(\boldsymbol{\gamma},\boldsymbol{\beta})$};
  \node[qbox] (qpu) at (6.6,0)
    {\textbf{Quantum processor}\\[1pt] prepares $|\psi(\boldsymbol{\theta})\rangle$\\ and measures it};
  \node[cbox, text width=3.6cm] (est) at (6.6,-2.7)
    {\textbf{Shots $\rightarrow$ estimate}\\[1pt] $S$ bitstrings $\mathbf{x}^{(1)},\dots,\mathbf{x}^{(S)}$\\ $\hat F(\boldsymbol{\theta})=\frac1S\sum_s C(\mathbf{x}^{(s)})$};
  \node[cbox] (out) at (0,-2.7)
    {\textbf{Output}\\[1pt] best bitstring seen,\\ then classical repair};

  \draw[->, line width=0.8pt] (opt) -- node[lab, above=1pt] {angles $\boldsymbol{\theta}$} (qpu);
  \draw[->, line width=0.8pt] (qpu) -- node[lab, right] {repeat $S$ times\\(shots)} (est);
  \draw[->, line width=0.8pt]
       (est.north west) -- node[lab, sloped, above=1pt] {objective estimate $\hat F(\boldsymbol{\theta})$} (opt.south east);
  \draw[->, line width=0.8pt, densely dashed, black!60] (opt) -- node[lab, left] {after\\convergence} (out);
\end{tikzpicture}

%% file: figures/fig2_pipeline.tex
\begin{tikzpicture}[
  >=Stealth,
  stage/.style={draw=black!65, line width=0.6pt, rounded corners=2.5pt,
                minimum height=1.05cm, text width=1.6cm, align=center,
                font=\sffamily\scriptsize\bfseries, inner sep=1.5pt},
  cstage/.style={stage, fill=black!6},
  qstage/.style={stage, fill=qblue!18, draw=qblue!80!black},
  ai/.style={draw=aiorange!85!black, fill=aiorange!10, rounded corners=2pt,
             text width=1.56cm, align=center, font=\sffamily\tiny,
             inner sep=2pt, minimum height=1.0cm,
             execute at begin node={\hyphenpenalty=10000\exhyphenpenalty=10000}},
  link/.style={densely dotted, aiorange!85!black, line width=0.6pt},
]
  \def\step{1.92}
  \foreach \i/\sty/\name/\aitag [count=\k from 0] in {
      1/cstage/{Problem\\instance}/{\textbf{solver selection}\\classifiers},
      2/cstage/{Formulation}/{\textbf{penalties, constraints}\\RL, surrogates},
      3/cstage/{Decom-\\position}/{\textbf{coarsening, partitions}\\GNN, multilevel},
      4/qstage/{Encoding\\\& circuit}/{\textbf{ansatz, operators}\\RL, QAS, LLM},
      5/qstage/{Solver\\configuration}/{\textbf{angles, queries}\\GNN, BO, meta-learning},
      6/qstage/{Hardware\\execution}/{\textbf{backend, shots}\\bandits, RL},
      7/cstage/{Repair \&\\reconstruction}/{\textbf{mitigation, repair}\\neural nets, RL}} {
    \node[\sty] (s\i) at (\k*\step,0) {\name};
    \node[ai, above=0.42cm of s\i] (a\i) {\aitag};
    \draw[link] (a\i) -- (s\i);
  }
  \foreach \a/\b in {1/2,2/3,3/4,4/5,5/6,6/7}
    \draw[->, line width=0.7pt, black!75] (s\a) -- (s\b);

  \node[draw=aiorange!85!black, fill=aiorange!22, rounded corners=3pt,
        fit=(a1)(a7), inner sep=0pt, minimum height=0.55cm,
        yshift=1.05cm, font=\sffamily\scriptsize, align=center] (pol) {};
  \node[font=\sffamily\scriptsize] at (pol)
       {\textbf{Learned solver policy} $\pi(a_t \mid s_t)$:
        coordinates several stage decisions (AI--quantum co-optimization)};
  \foreach \i in {1,...,7}
    \draw[->, aiorange!85!black, line width=0.5pt] (pol.south -| a\i) -- (a\i.north);

  \draw[->, line width=0.8pt, qblue!80!black, rounded corners=6pt]
       (s7.south) -- ++(0,-0.55) -| (s2.south)
       node[pos=0.25, below, font=\sffamily\scriptsize\itshape, text=black!80]
       {execution feedback (diagnostics $d_t$, remaining budget $r_t$) can trigger
        reformulation, repair, or rerouting};

  \node[cstage, minimum height=0.3cm, text width=0.3cm, font=\tiny] (lg1)
       at ([xshift=0.25cm,yshift=-1.45cm]s1.south west) {};
  \node[right=2pt of lg1, font=\sffamily\scriptsize] (lt1) {classical stage};
  \node[qstage, minimum height=0.3cm, text width=0.3cm, right=10pt of lt1] (lg2) {};
  \node[right=2pt of lg2, font=\sffamily\scriptsize] (lt2) {quantum-facing stage};
  \node[ai, minimum height=0.3cm, text width=0.3cm, right=10pt of lt2] (lg3) {};
  \node[right=2pt of lg3, font=\sffamily\scriptsize] {AI intervention (decision \textbf{learned}, typical methods)};
\end{tikzpicture}

%% file: figures/fig3_autonomy_evidence.tex
\begin{tikzpicture}[
  lbl/.style={font=\sffamily\scriptsize, text=black!85, inner sep=1pt},
  mrange/.style={line width=2.4pt, line cap=round, opacity=0.35},
]
\begin{axis}[
  scale only axis, width=\linewidth-3.7cm, height=6.2cm,
  xmin=0.55, xmax=4.45, ymin=0.6, ymax=4.55,
  xtick={1,2,3,4},
  xticklabels={{Parameter\\learning},{Component\\design},{Policy\\control},{Automated\\policy search}},
  xticklabel style={align=center, font=\sffamily\footnotesize},
  ytick={1,2,3,4},
  yticklabels={{\textbf{M1} simulation},{\textbf{M2} generalization},
               {\textbf{M3} hardware component},{\textbf{M4} end-to-end hardware}},
  yticklabel style={font=\sffamily\footnotesize, align=right, text width=2.55cm},
  xlabel={Scope of learned solver control $\longrightarrow$},
  ylabel={Strongest quantum evidence},
  label style={font=\sffamily\footnotesize},
  grid=major, grid style={black!10, line width=0.4pt},
  axis line style={black!60}, tick style={black!60},
  major tick length=2pt,
  legend style={font=\sffamily\scriptsize, at={(0.985,0.03)}, anchor=south east,
                draw=black!25, fill=white, fill opacity=0.92, text opacity=1,
                row sep=-1pt, inner sep=2pt},
  legend cell align=left,
  clip=false,
]
\fill[black!4] (axis cs:0.55,4.25) rectangle (axis cs:4.45,4.55);
\node[font=\sffamily\scriptsize\itshape, text=black!55, anchor=east]
     at (axis cs:4.42,4.40) {\textbf{M5} practical quantum advantage: not reached in the screened corpus};

\draw[mrange, coopt]  (axis cs:3.55,3) -- (axis cs:3.55,4);   
\draw[mrange, coopt]  (axis cs:4.00,2) -- (axis cs:4.00,3);   
\draw[mrange, qorange](axis cs:2.12,3) -- (axis cs:2.12,4);   

\addplot[only marks, mark=*, mark size=2.6pt, color=qblue, fill=qblue]
  coordinates {(1.00,2) (1.18,3) (1.42,3) (1.68,3) (2.58,2)};
\addlegendentry{AI-for-QO}
\addplot[only marks, mark=square*, mark size=2.4pt, color=qorange, fill=qorange]
  coordinates {(2.12,3.5) (2.00,2) (2.28,2)};
\addlegendentry{Q-for-AI}
\addplot[only marks, mark=triangle*, mark size=3.2pt, color=coopt, fill=coopt]
  coordinates {(2.90,3) (3.18,3) (3.55,3.5) (4.00,2.5)};
\addlegendentry{AI--quantum co-optimization}

\node[lbl, anchor=north]      at (axis cs:1.00,1.90) {Jain et al.};
\node[lbl, anchor=north east] at (axis cs:1.20,2.92) {Sack \& Egger};
\node[lbl, anchor=south]      at (axis cs:1.36,3.08) {O'Leary et al.};
\node[lbl, anchor=north west] at (axis cs:1.60,2.92) {Cheng et al.};
\node[lbl, anchor=east]       at (axis cs:2.07,3.55) {Correll et al.};
\node[lbl, anchor=north east] at (axis cs:2.02,1.93) {Skolik et al.};
\node[lbl, anchor=north west] at (axis cs:2.26,1.93) {Kruse et al.};
\node[lbl, anchor=south]      at (axis cs:2.58,2.08) {QAOA-GPT};
\node[lbl, anchor=south]      at (axis cs:2.86,3.08) {Rezk \& Gora};
\node[lbl, anchor=north]      at (axis cs:3.18,2.92) {Lagrangian CVRP};
\node[lbl, anchor=west]       at (axis cs:3.60,3.5)  {Q-ALNS};
\node[lbl, anchor=east]       at (axis cs:3.95,2.5)  {AutoQResearch};
\end{axis}
\end{tikzpicture}

%% file: figures/fig4_maturity_by_paradigm.tex
\begin{tikzpicture}
\begin{axis}[
  scale only axis, width=\linewidth-4.3cm, height=3.3cm,
  xbar stacked, bar width=0.46cm,
  xmin=0, xmax=100, ymin=0.45, ymax=3.55,
  ytick={1,2,3},
  yticklabels={{AI--quantum co-opt.\ ($n{=}10$)},{Q-for-AI ($n{=}15$)},{AI-for-QO ($n{=}27$)}},
  yticklabel style={font=\sffamily\footnotesize},
  xlabel={Share of coded studies in paradigm (\%)},
  label style={font=\sffamily\footnotesize},
  xticklabel style={font=\sffamily\footnotesize},
  axis line style={black!60}, tick style={black!60}, major tick length=2pt,
  y axis line style={draw=none}, ytick style={draw=none},
  xmajorgrids, grid style={black!10},
  legend style={font=\sffamily\scriptsize, at={(0.5,1.03)}, anchor=south,
                legend columns=6, draw=none, /tikz/every even column/.append style={column sep=6pt}},
  legend image code/.code={\draw[#1, draw=none] (0cm,-0.09cm) rectangle (0.32cm,0.09cm);},
  nodes near coords,
  nodes near coords align={center},
  every node near coord/.append style={font=\sffamily\scriptsize},
  point meta=explicit symbolic,
  enlarge y limits=false,
]
\addplot[fill=mat1, draw=white, every node near coord/.append style={text=black}]
  coordinates {(0,1) [] (13.333,2) [2] (3.704,3) [1]};
\addplot[fill=mat2, draw=white, every node near coord/.append style={text=black}]
  coordinates {(10,1) [1] (13.333,2) [2] (11.111,3) [3]};
\addplot[fill=mat3, draw=white, every node near coord/.append style={text=black}]
  coordinates {(50,1) [5] (60,2) [9] (59.259,3) [16]};
\addplot[fill=mat4, draw=white, every node near coord/.append style={text=white}]
  coordinates {(10,1) [1] (0,2) [] (7.407,3) [2]};
\addplot[fill=mat5, draw=white, every node near coord/.append style={text=white}]
  coordinates {(20,1) [2] (6.667,2) [1] (18.519,3) [5]};
\addplot[fill=mat6, draw=white, every node near coord/.append style={text=white}]
  coordinates {(10,1) [1] (6.667,2) [1] (0,3) []};
\legend{M1, M1--M2, M2, M2--M3, M3, M3--M4}
\end{axis}
\end{tikzpicture}